\documentclass[twocolumn,twocolumn,tighten,times]{aastex631}

\newcommand{\mum}{$\mu$m}

\usepackage{longtable}
\usepackage{amsmath} 
\usepackage{scrextend}
\usepackage{float}

\newcommand{\RNum}[1]
{\uppercase\expandafter{\romannumeral #1\relax}}

\providecommand{\sorthelp}[1]{}
\usepackage{array}
\newcolumntype{P}[1]{>{\begin{minipage}[t]{#1}\raggedright\arraybackslash}l<{\end{minipage}}}
\graphicspath{{./}{}}
\begin{document}

\title{Far-infrared Polarization Properties of Nearby Star-forming Regions: A New Compendium of SOFIA/HAWC+ Observations}

\author[0009-0006-4830-163X]{Kaitlyn Karpovich}
\affiliation{Department of Physics, Stanford University, Stanford, CA 94305, USA}
\affiliation{Kavli Institute for Particle Astrophysics \& Cosmology (KIPAC), Stanford University, Stanford, CA 94305, USA}

\author[0000-0002-7633-3376]{Susan E. Clark}
\affiliation{Department of Physics, Stanford University, Stanford, CA 94305, USA}
\affiliation{Kavli Institute for Particle Astrophysics \& Cosmology (KIPAC), Stanford University, Stanford, CA 94305, USA}

\author[0000-0001-5357-6538]{Enrique Lopez-Rodriguez}
\affiliation{Department of Physics \& Astronomy, University of South Carolina, Columbia, SC 29208, USA}

\begin{abstract}

We present a comprehensive polarimetric study of 26 nearby molecular clouds in four far-infrared bands (53 $\mu$m~to 214~$\mu$m) using 52 archival SOFIA/HAWC+ datasets{, over half of which are unpublished}. Far-infrared dust polarization observations probe the plane-of-sky magnetic field. To investigate scale-dependent trends, we group the molecular clouds by distance and analyze the data at common angular ($25''$) and common physical (0.052 pc and 0.32 pc) resolutions. The two shorter wavelengths are more impacted by smoothing, exhibiting a larger decrease in percent polarization. We analyze the polarization spectrum --- the polarization fraction as a function of wavelength --- and find that it depends more strongly on column density than dust temperature. We find a ``falling" spectrum at the 0.052 pc resolution, but find a ``flat" spectrum at the 0.32 pc resolution, suggesting that resolution plays an important role in the observed polarization spectra. We propose that warm dust grain emission in small-scale structures ($\lesssim$~0.1 pc) traces different magnetic field geometries only resolved in our {near} regime data. There is no preferred magnetic field orientation across our data, which suggests that the magnetic field in our $\sim$~parsec scale regions is decoupled from the large-scale field that is primarily parallel to the Galactic plane. The relationship between percent polarization and column density varies between clouds, but the correlation between percent polarization and {polarization angle} dispersion is consistent across regions. This compendium 
highlights the value of observing at multiple far-infrared wavelengths and 
enables additional population-level studies of magnetic fields and dust across star-forming environments. 

\end{abstract}


\section{Introduction} \label{sec:intro}

Star formation typically occurs in the densest parts of molecular clouds; however, it has been known for decades that star formation occurs at a rate lower than expected from the masses of molecular clouds \citep[e.g.,][]{Zuckerman1974, Krumholz2007}. Proposed explanations for this often invoke magnetic fields and/or turbulence to explain the low rate of star formation \citep[e.g.,][]{Mouschovias1976, MacLow2004, Federrath2012}. Understanding the interplay between magnetic fields, turbulence, and gravity is crucial for understanding the complete picture of star formation efficiency. 

Methods to observe the plane-of-sky magnetic field orientation in molecular clouds include optical/near-infrared absorption polarization {\citep{Hall1949, Hiltner1949}} and far-infrared/submillimeter emission polarization \citep{Hildebrand2000}. Far-infrared wavelengths, typically defined as $\lambda \sim 50-300$ \mum, are ideal for studying magnetic fields through polarized dust emission, because thermal dust emission peaks in the far-infrared, where both warm ($\sim 30-60$ K) and cold ($\sim 10-30$ K) dust populations contribute significantly to the observed flux. Plane-of-sky magnetic fields can be observed via the polarized radiation from elongated dust grains that align their short axes with the magnetic field. The leading explanation for how grains become aligned involves Radiative Alignment Torques (RAT), in which {anisotropic radiation fields cause paramagnetic grains to rotate suprathermally.} {This rotation can} induce a magnetic moment due to the Barnett {e}ffect {that allows the grains to align with the magnetic field} \citep{Dolginov1976, Draine1997,Lazarian2007}.

Multiwavelength observations of polarized dust emission are particularly valuable. Dust polarization spectra in the far-infrared and submillimeter, which quantify how the percent polarization changes as a function of wavelength, are important for understanding the physical properties of interstellar dust grains \citep[e.g.,][]{Draine2021} and grain alignment \citep[e.g.,][]{Bethell2007, Lee2024,Tram2024}. The submillimeter spectra have been observed to be relatively flat \citep[e.g.,][]{Ashton2018,Henseley2021}. However, some observations have found that the far-infrared portion of the spectrum in molecular clouds can have a ``V" shape, with the polarization fraction decreasing with wavelength (falling) until about 300 \mum~, and then increasing (rising) at longer wavelengths \citep{Vaillancourt2002}. This complex spectral dependence can be caused by having multiple cloud components with different temperatures along the line of sight. Studying the polarization spectrum as a function of cloud properties such as temperature and density can be used to understand the three-dimensional geometry of clouds \citep{Lee2024, Seifried2025}.

There have been studies in the far-infrared wavelength range that explore the polarization properties in molecular clouds using instruments on the Kuiper Airborne Observatory (KAO) and the Caltech Submilimeter Observatory (CSO) \citep[e.g.,][]{Hildebrand_1999, Vaillancourt2002, Stephens2011}. The Planck satellite has also been used to study polarized Galactic emission at 353 GHz ($\sim 850$ \mum), revealing the large-scale structure of magnetic fields in the diffuse and moderately dense interstellar medium \citep[e.g.,][]{PlanckXII, PlanckXIX}. However, Planck's angular resolution is $4.94'$ at 353 GHz and to get high enough signal to noise the data are often smoothed to a resolution of $10'$ or worse \citep[e.g.,][]{PlanckXII, PlanckXXXV}. 

The most recent far-infrared polarimeter was the High-resolution Airborne Wideband Camera-Plus \citep[HAWC+;][]{Harper2018} on the Stratospheric Observatory For Infrared Astronomy (SOFIA). HAWC+ observed linear polarization in the far-infrared at wavelengths of 53, 89, 154, and 214 \mum. The resolution of the HAWC+ data ($\leq18.2''$) allows the study of sub-parsec structure in molecular clouds out to a distance of $\sim$10 kpc. Compared to instruments like Planck, HAWC+ is able to resolve higher-density structures, with our work finding column densities in the range $N_{H_2} \sim 10^{22}-10^{24}$ cm$^{-2}$. These dense regions may contain substantial substructure, including $\lesssim0.1$ pc hot molecular cores with temperatures $\sim100 ~\text{K}$ \citep{Cesaroni2005, Andre2010}; filaments with widths of order $\sim0.1$ pc \citep{Arzoumanian2011, Hennemann2012, Schisano2014}; and denser sub-filaments \citep{FernandezLopez2014, Hacar:2013}. 

Currently, there are no studies using a large compilation of HAWC+ datasets over a wide variety of molecular clouds. Many works are comprehensive studies of the properties of a single cloud at a time \citep[e.g.,][]{Santos2019, Chuss2019, Zielinski2022, Pillai2020, Bij2024, Hoang2022, Sarkar2025, Seo2021} or multiple components of one cloud complex \citep[e.g.,][]{Lee2021, Barnes2025}. There are some works that look at multiple diverse clouds like the three clouds studied in \citet{Cox2025} and the ten ``bones" studied in \citet{Stephens2025} and \citet{Coude2026}. However, the full potential of archival HAWC+ observations of molecular clouds has yet to be realized.

To understand the properties of magnetic fields and polarization broadly across diverse environments, we compile a large sample of 52 archival SOFIA/HAWC+ datasets, over half of which are currently unpublished. Our data include 26 star-forming regions spanning distances from $130$ to $2620$ pc. We also use data at wavelengths of 70, 100, 160, 250, and 350 \mum~from the Herschel Space Observatory to derive temperature and column density maps. We analyze the data at common angular and common physical resolutions to determine the effects of resolution and distance on polarization properties. We choose two physical resolutions of 0.052 pc and 0.32 pc, with the smaller scale able to resolve many small cores and filaments.

This paper is organized as follows: we first describe the far-infrared data products and processing from the HAWC+ instrument on SOFIA and the Herschel Space Observatory  (\autoref{sec:obs}); we discuss our selected compilation and detail how we separate them into regimes by distance (\autoref{sec:sample}); we describe how we compute the derived quantities of column density, temperature, and angular dispersion we need for our analyses (\autoref{sec:derived_prop}); and we detail our analyses and results in the context of past literature (\autoref{sec:results}). Finally, we present a summary of our key results in \autoref{sec:conclusion}.

\section{Observations} \label{sec:obs}

\subsection{HAWC+ Data}\label{subsec:HAWC}
We use 52 archival imaging polarimetric datasets from the HAWC+ \citep{Harper2018} instrument on SOFIA. HAWC+ observed in four wavelength bands centered on 53 \mum~(bandwidth $\Delta\lambda=8.7$ \mum, Band A), 89 \mum~($\Delta\lambda=17$ \mum, Band C), 154 \mum~($\Delta\lambda=34$ \mum, Band D), 214 \mum~($\Delta\lambda=44$ \mum, Band E) with resolutions of $4.85''$, $7.8''$, $13.6''$, and $18.2''$, respectively. Our collection consists of nine Band A, twelve Band C, fifteen Band D, and sixteen Band E maps, and we note the available bands for each region in \autoref{tab:regions}. We have only chosen datasets that were observed using the chop-nod mode of HAWC+ because the on-the-fly mapping (OTFMAP) mode requires a different reduction algorithm \citep{LR2022}. The on-the-fly mapping mode also does not effectively recover large-scale features in the data \citep{LR2022}, which we need for our analyses at common angular and physical resolutions.

Data were reduced using the SOFIA Redux pipeline\footnote{\url{https://irsa.ipac.caltech.edu/data/SOFIA/docs/data/data-pipelines/}\label{HAWC}}. Some of the data on the archive needed to be re-reduced, either because the only level 4 products available were produced with an outdated version of the pipeline or because the level 4 products did not include all of the available observations for the region. The final data used are a combination of level 4 data downloaded from the SOFIA/HAWC+ archive {\citep{https://doi.org/10.26131/irsa636}} that used version 3.2.0 of the pipeline, data we reduced using version 2.7.0 of the pipeline, and data reduced for \citet{Chuss2019} using version 1.3.0beta3 of the pipeline. {The HAWC+ data used in this work are made publicly available on Harvard Dataverse\footnote{\url{https://doi.org/10.7910/DVN/SOKL2G}\label{Dataverse}}}. There were no substantial differences in the chop-nod data reduction between versions 3.2.0 and 2.7.0. The updates between these versions focused mainly on the data reduction for the OTFMAP observations and minor bug fixes\footref{HAWC}. We confirm that the versions produce similar results by re-reducing data with version 2.7.0 and comparing the results with data on the archive reduced with version 3.2.0. Even though \citet{Chuss2019} used an older version of the pipeline, we still use their data because their reduction and processing takes into account potential background contamination from off-axis chops. \autoref{tab:obs_table} denotes which pipeline was used for each dataset. We reduce the data with the default parameters and remove files that were flagged by the archive quality assurance notes for data quality issues. We also check the individual files produced by the ``Calibrate Flux" step in the pipeline to look for and remove anomalies that impact the final merged maps. The headers of the final FITS files denote the raw files used.

Bands A and C from the W3 region included data observed in the ``Standard Flux" mode in addition to the usual mode. To reduce these data into a single map, we force the pipeline to reduce all the data in \textsc{nodpol} mode instead of \textsc{nodpol\_std} by replacing the steps used in the pipeline for the \textsc{nodpol\_std} mode with the step of the \textsc{nodpol} mode. 

The final data products are maps of the linear Stokes parameters ($I$, $Q$, and $U$) from which maps of the {polarization fraction} ($p$) and the polarization angle ($\phi$) are computed, along with statistical uncertainties.
\begin{equation}
\label{eq:p}
    p = \frac{\sqrt{Q^2+U^2}}{I}
\end{equation}
\begin{equation}
    \phi=\frac{1}{2}\text{arctan2}\left(U,Q\right)
\end{equation}
The uncertainties on $p$ ($\sigma_p$) and $\phi$ ($\sigma_\phi$) are propagated as
\begin{equation}
    \sigma_p=\frac{1}{I}\sqrt{\frac{(\sigma_QQ)^2+(\sigma_UU^2)}{Q^2+U^2}+\frac{\sigma_I^2}{I^2}(Q^2+U^2)}
\end{equation}
\begin{equation}
    \sigma_\phi=\frac{1}{2}\frac{1}{Q^2+U^2}\sqrt{(\sigma_UQ)^2+(\sigma_QU)^2}.
\end{equation}
Our polarization orientation is defined in the IAU convention, where positive $\phi$ is measured counterclockwise from North when looking toward the source. Due to its strictly positive nature, $p$ is strongly biased at low signal-to-noise \citep{Plaszczynski2014}. We debias $p$ using its statistical uncertainty as follows:
\begin{equation}
    p_{debiased}=\sqrt{p^2-\sigma_p^2}.
\end{equation}
We use $p_{debiased}$ for all of our analyses and will hereafter refer to it as $p$. We then apply signal-to-noise cuts to only include data that satisfy $I/\sigma_I\geq200$ and $p/\sigma_p\geq3$.

\subsection{Herschel data}\label{subsec:herschel}
We also use archival total intensity data from the Photodetector Array Camera and Spectrometer \citep[PACS;][]{PACS} and Spectral and Photometric Imaging REceiver \citep[SPIRE;][]{SPIRE} instruments on the Herschel Space Telescope to fit modified blackbody functions in each region. PACS observed at 70 \mum, 100 \mum, and 160 \mum. The angular resolution at each wavelength is lengthened along the scan direction with a magnitude of elongation that depends on the scan speed and whether the observation was taken in the PACS/SPIRE parallel mode, which is modeled as a two-dimensional Gaussian as described in \citet{PACS}. When smoothing the PACS data, we use the geometric mean of the two Gaussian widths for each dimension as the native resolution of the data. For the 70 \mum~ and 160 \mum~ data, we have a combination of data sets taken at 20 arcsec/sec and 60 arcsec/sec in parallel mode. All of the 100 \mum~ data were observed at 20 arcsec/sec. This results in native resolutions of $5.6''$ and $8.4''$ (70 \mum), $6.8''$ (100 \mum), and $11.2''$ and $13.4''$ (160 \mum). We also use 250 \mum~and 350 \mum~SPIRE data, which have angular resolutions of $18.2''$ and $25''$. We smooth to the resolution of the SPIRE 350 \mum~data as they are the lowest resolution data between both Herschel and HAWC+. We adopt uncertainties of 20\% for PACS and 10\% for SPIRE \citep{PACS, SPIRE}. Not every region had observations at all five wavelengths and we note which wavelengths were used in \autoref{tab:regions}. 

The thermal dust spectral energy distribution (SED) is not constant across the wide Herschel filters, and is instead characterized by a slope that depends on the source's spectral index and temperature. We thus apply color correction factors to both the PACS and SPIRE data. The color corrections are tabulated in \citet{PACS_color} and the SPIRE handbook\footnote[2]{\url{http://herschel.esac.esa.int/Docs/SPIRE/spire_handbook.pdf}} as a function of blackbody temperature and spectral index. Preliminary fits of the intensity data without color corrections showed a general temperature range between 15 K and 60 K and a spectral index range between 1 and 2. To determine the color correction factors, we average the values in the tables of \citet{PACS_color} and the SPIRE handbook for this range of temperatures and spectral indices. This results in multiplicative correction values of 0.843, 0.996, 0.986, 0.991, and 0.981 for the 70 \mum, 100 \mum, 160 \mum, 250 \mum, and 350 \mum~ bands, respectively. After applying these color correction factors, the resulting temperatures and spectral indices still fall within the original ranges.

\begin{longtable*}{ 
P{1.7cm}|P{0.95cm}|P{.99cm}|P{2.1cm}|P{2.8cm}|P{1.8cm}|P{3.45cm}|P{1.5cm}}
\label{tab:regions}
\textbf{Region} & \textbf{RA ($^{\circ}$)} & \textbf{Dec ($^{\circ}$)} &\textbf{HAWC+ Bands}&\textbf{Herschel $\lambda$ (\mum)}&\textbf{Distance (pc)}&\textbf{Distance Reference} &\textbf{$\beta$}\\
\endfirsthead
\toprule
L1527           &   69.978  &   26.046& D  &  70, 160, 250, 350  &   130  &  \citet{Roccatagliata2020}  & 1.00$\pm$ 0.01\\
L1688           &  246.628  &  -24.394  &  A, C, D &70, 100,160, 250, 350  &   139  &  \citet{Zucker2020} & 1.50$\pm$0.02        \\
$\rho$ Oph E/F  &  246.782  &  -24.626  & D & 70, 100,160, 250, 350  &   139  &  \citet{Zucker2020} & 1.00$\pm$0.001        \\
L1689N          &  248.094  &  -24.476  &C, D&  100, 160, 250, 350      &   147  &  \citet{Ortiz-Leon2017} &1.10$\pm$0.04    \\
\hline \hline
NGC 1333         &   52.291  &   31.369  &  C, E &70, 100,160, 250, 350  &   299  &  \citet{Zucker2018} &1.00$\pm$0.001\\
OMC1            &   83.803   &   -5.371  &  A, C, D, E &70, 100,160, 250, 350  &   388  &  \citet{Kounkel2017} &2.02$\pm$0.01      \\
OMC2/3          &   83.809  &   -4.983  &  D, E &70, 100,160, 250, 350  &   388  &  \citet{Kounkel2017} &1.43$\pm$0.01       \\
NGC 2071        &   86.768  &    0.361  &  E &70, 100,160, 250, 350  &   388  &  \citet{Fedriani2023} &1.00$\pm$0.004      \\
NGC 2068        &   86.651  &    0.042  &  E&70, 100,160, 250, 350  &   388  &  \citet{Kounkel2017} &1.64$\pm$0.01       \\
Serpens       &  277.506  &   -2.054  &E&  70, 100,160, 250, 350  &   432  &  \citet{Ortiz-Leon2018} &1.00$\pm$0.0002    \\
\hline
Vela C          &  134.839  &  -43.767  &C, E&  70, 160, 250, 350      &   933  &  \citet{Fissel2019} &1.66$\pm$0.01        \\
S106            &  306.855  &   37.388  &  A, E&70, 160, 250, 350      &  1091  &  \citet{Zucker2020} &1.68$\pm$0.01        \\
NGC 6334         &  260.078  &  -35.914  &  A, C, D, E& 70, 160, 250, 350      &  1350  &  \citet{Wu2014}   &1.96$\pm$0.01         \\
DR21            &  309.754  &   42.378  &A, E&  70, 160, 250, 350      &  1400  &  \citet{Rygl2012} &1.81$\pm$0.02          \\
G34.43          &  283.340  &    1.421  &E&  70, 160, 250, 350      &  1560  &  \citet{Kurayama2011}  & 1.24$\pm$0.02     \\
M17             &  275.097   &  -16.211  &A, C, D, E&  70, 100,160, 250, 350  &  1600  &  \citet{Zucker2020} &1.54$\pm$0.01        \\
G14.2N          &  274.551  &  -16.837  &D&  70, 160, 250, 350      &  1600  &  \citet{Diaz-Marquez2023} &1.18$\pm$0.02  \\
G14.2S          &  274.551  &  -16.957  &D&  70, 160, 250, 350      &  1600  &  \citet{Diaz-Marquez2023} &1.24$\pm$ 0.02 \\
M16             &  274.719  &  -13.833  &C, D&  70, 160, 250, 350      &  1700  &  \citet{Zucker2020} &1.60$\pm$0.02        \\
G351.77    &  261.662  &  -36.121  &E&  70, 160, 250, 350      &  2000  &  \citet{Reyes-Reyes2024} &1.24$\pm$0.01   \\
W3OH            &   36.764  &   61.874  &A, C, D, E&  70, 160, 250, 350      &  2000  &  \citet{Navarete2019}  &1.18$\pm$0.01     \\
W3              &   36.417  &   62.104   &A, C, D, E&  70, 160, 250, 350      &  2300  &  \citet{Navarete2019} &1.95$\pm$0.01      \\
W33             &  273.558  &  -17.928  &A, C, E&  70, 160, 250, 350      &  2400  &  \citet{Immer2013}  &1.83$\pm$0.02        \\
BYF77           &  159.437  &  -58.772  &D&  70, 160, 250, 350      &  2500  &  \citet{Vazzano2014} &1.74$\pm$0.01       \\
BYF73           &  159.634  &  -58.311  &D&  70, 160, 250, 350      &  2500  &  \citet{Barnes2010}  &1.41$\pm$0.02       \\
Keyhole         &  161.220  &  -59.654  &C&  70, 160, 250, 350      &  2620  &  \citet{Kuhn2019} &1.43$\pm$0.01          \\
\caption{A table of every region included in our compilation, with their location in equatorial coordinates, the available Herschel wavelengths, their distances and reference, and the dust emissivity index determined in \autoref{sec:derived_prop}. Every region above the single solid line is in the {near} regime (130-432 pc) and every region below the double solid line is in the far regime (299-2620 pc).}
\end{longtable*}

\section{Compilation}
\label{sec:sample}

\begin{figure*}
    
    \centering
    \includegraphics[width=\textwidth]{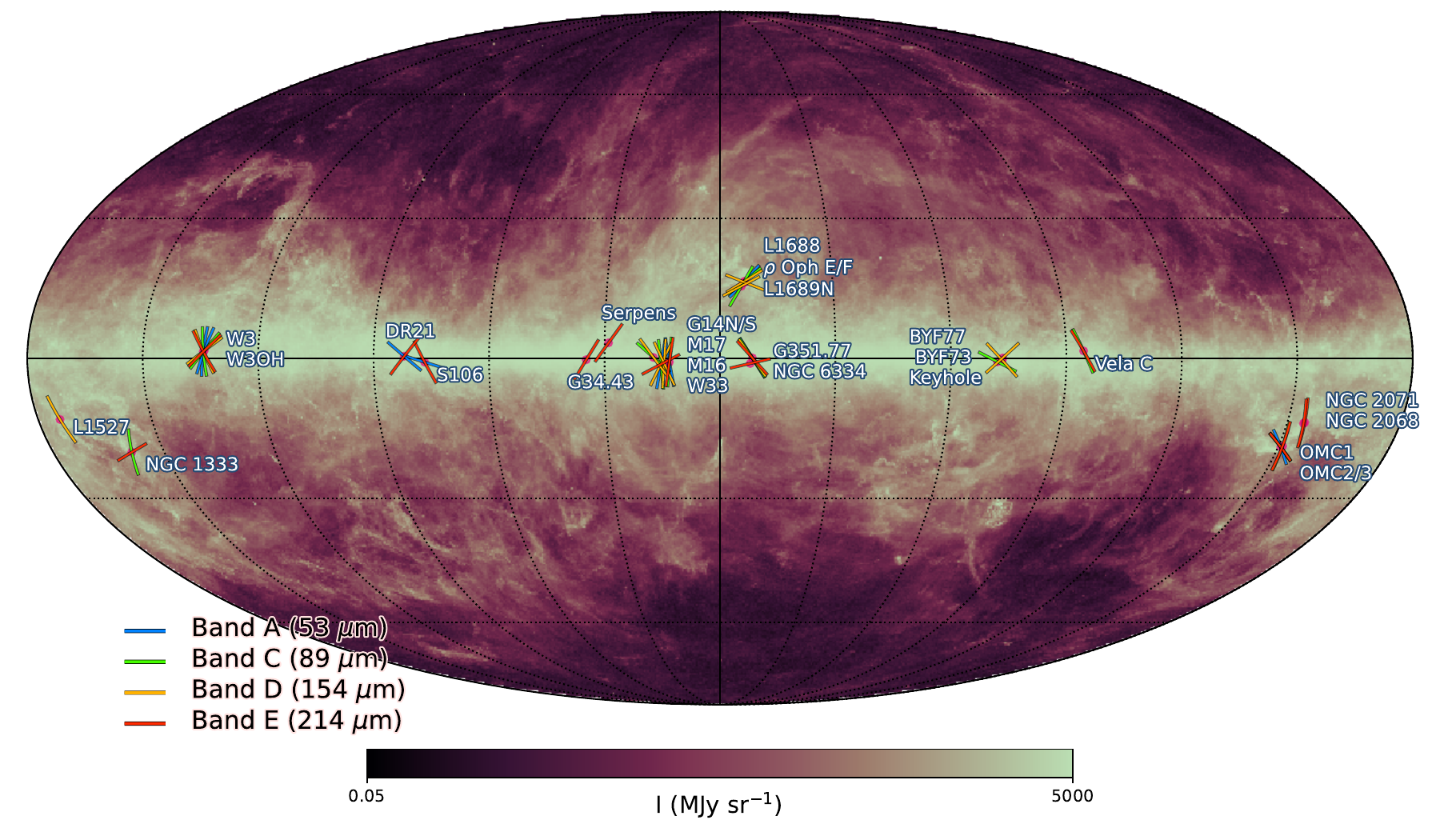}
    \caption{Sky distribution of the 26 regions used in this paper. Each region is marked with a pink dot on a galactic projection full-sky map of the Planck 857 GHz data with histogram equalized color mapping. The vectors represent the average magnetic field orientation for each region as a function of wavelength (colors shown in legend).}
    \label{fig:fullsky}
\end{figure*}

\begin{figure*}
    
    \centering
    \includegraphics[width=\textwidth]{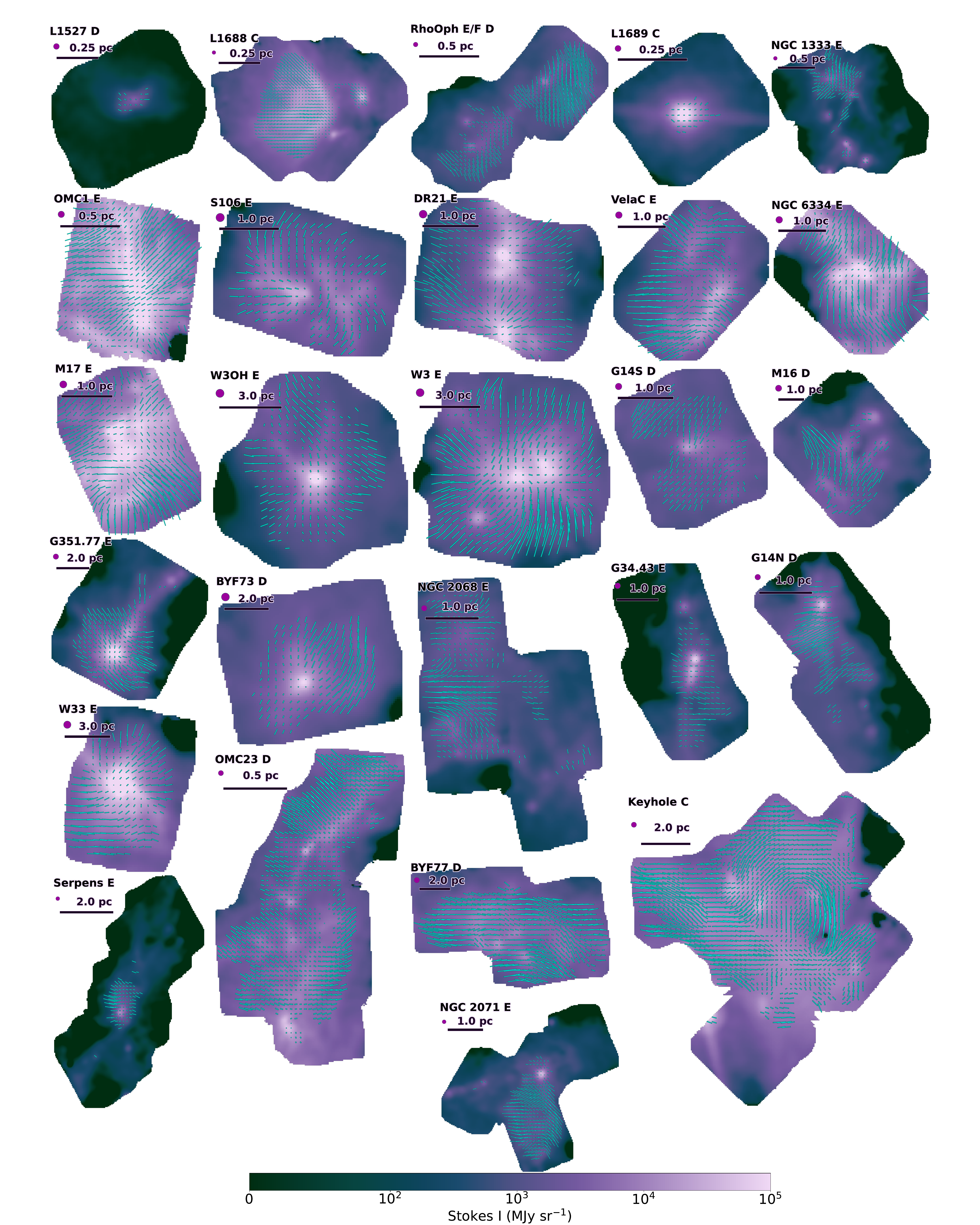}
    \caption{A representative HAWC+ dataset from each of the 26 regions. Each region is labeled with its name and a letter denoting the HAWC+ band shown. The purple circle shows the beam size. The vectors are the magnetic field pseudovectors and their length corresponds to the percent polarization. The pseudovectors are sampled so there is one vector per beam. The scale bar denotes the physical size in parsecs calculated from the distances in \autoref{tab:regions}.}
    \label{fig:fam_port}
\end{figure*}

\begin{figure*}
    \centering
    \includegraphics[width=\textwidth]{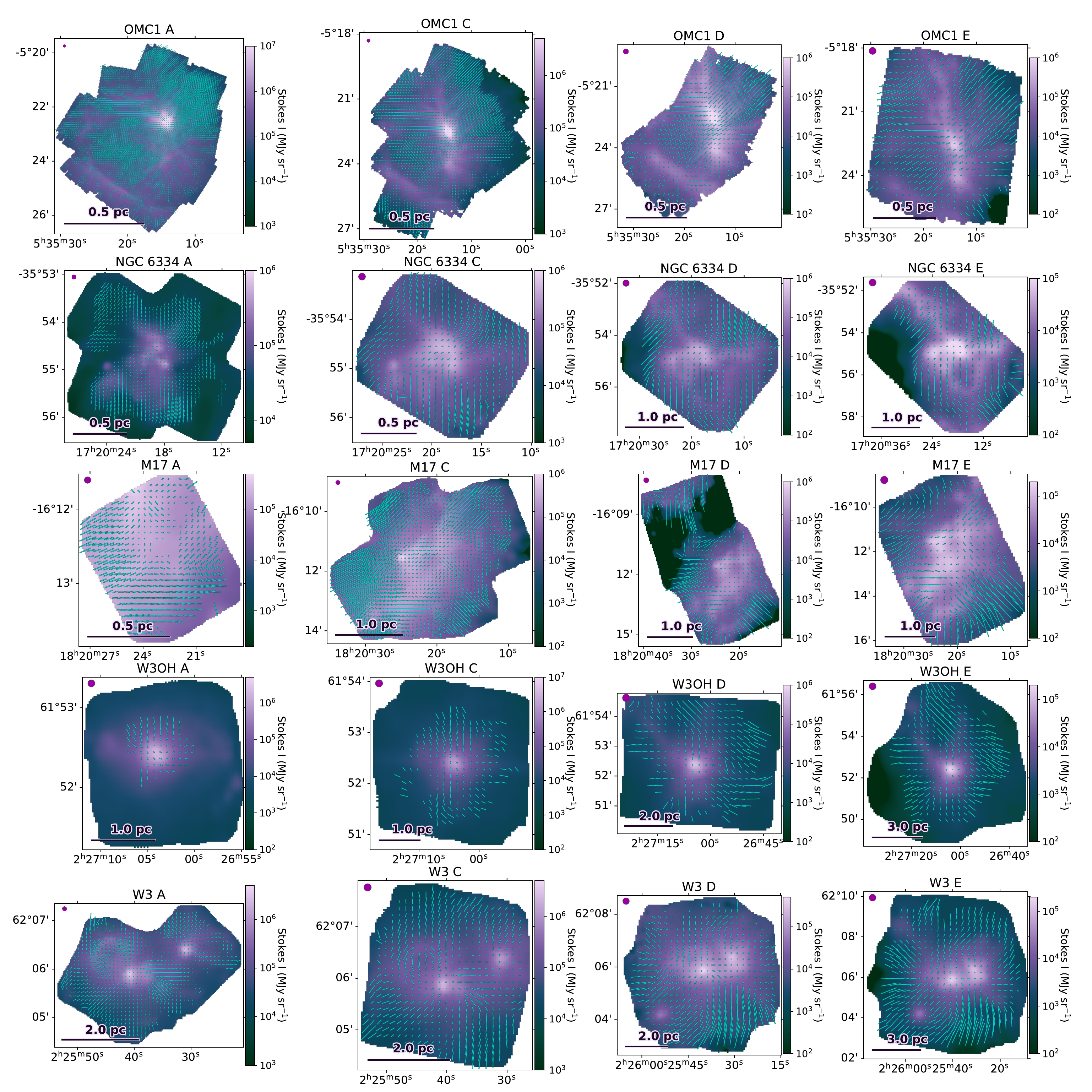}
    \caption{Same as \autoref{fig:fam_port} for the regions that we have coverage in all four HAWC+ bands.}
    \label{fig:four_band}
\end{figure*}

Our 52 dataset compilation comprises of data from 26 nearby star-forming regions from the SOFIA/HAWC+ archive that were observed with the chop-nod method and resulted in significant polarization detections. Our data cover a range of objects such as molecular clouds, filamentary clouds, infrared dark clouds (IRDCs), prestellar cores, and protostellar cores. HAWC+ observations cover some of the highest $H_2$ column density portions of the clouds ($\sim10^{22}-10^{23} \text{ cm}^{-2}$). While we may not directly measure emission from HII regions, young stellar objects (YSOs), and MASERs within these larger objects, they often dictate the observation centers and target names. \autoref{fig:fullsky} shows the location of every region in our compilation overlaid on an all-sky map of the Planck 857 GHz emission data \citep{PlanckIII}. In \autoref{tab:regions}, we list every region used in this analysis with its equatorial coordinates and a literature estimate of its distance from the Sun. We include a citation for every distance we use in the sixth column. In \autoref{fig:fam_port} we show an example map from each region using the band with approximately largest spatial coverage. While we do not have uniform wavelength coverage for every region, \autoref{fig:four_band} shows our regions that do have observations in all four HAWC+ bands. The rest of our HAWC+ data not included in \autoref{fig:fam_port} or \autoref{fig:four_band} are shown in \autoref{fig:remaining_data}.

For some of our analysis we smooth subsets of our data to a common physical resolution. We smooth the data by convolving with a Gaussian kernel. The map is then rebinned so that the width of four pixels equals the final beam full width at half maximum ($FWHM$) (i.e., Nyquist sampling). We propagate the uncertainties on the linear Stokes parameters by dividing the unsmoothed uncertainty at each pixel by $\sqrt{n}$ where $n=FWHM_f^2/FWHM_i^2$, the ratio of the final to the initial $FWHM$. The farthest region in a given distance bin determines how coarsely we must smooth the nearest regions. To ensure we have enough independent measurements ($>3$) in each region after smoothing to a common physical resolution in the nearest regions, we define two subsets of data: a {near} regime ($D<432$ pc) and a far regime ($299<D<2620$ pc). Both regimes include the regions with distances $299<D<432$, because these regions have more than 3 independent measurements when smoothed to the constant physical resolution of the far regime. The minimum number of independent measurements in a region in both regimes is 3, but most regions greatly exceed our minimum threshold. There are 10 regions in the {near} regime and 22 regions in the far regime, with 6 regions included in both regimes. 

The data are smoothed to the physical resolution of the Herschel 350 \mum~data resolution ($25''$) at 432 pc and 2620 pc, which results in resolutions of 0.052 pc and 0.32 pc, respectively. For our analyses, it is important to note that multiple regions in the {near} regime are near the 432 pc distance cut, meaning they are not smoothed with a large kernel in the common physical resolution analyses. Because of this, we do not necessarily expect large changes when we smooth the {near} regime data. We also note that one of the regions in the overlap is OMC1, which can contribute a large number of data points in all four bands at the common angular resolution due its large spatial coverage. The trends we see in our compilation remain when we remove OMC1, so we do not believe OMC1 is significantly biasing our results.

\section{Derived Properties}
\label{sec:derived_prop}

\begin{figure*}
    \centering
    \includegraphics[width=\textwidth]{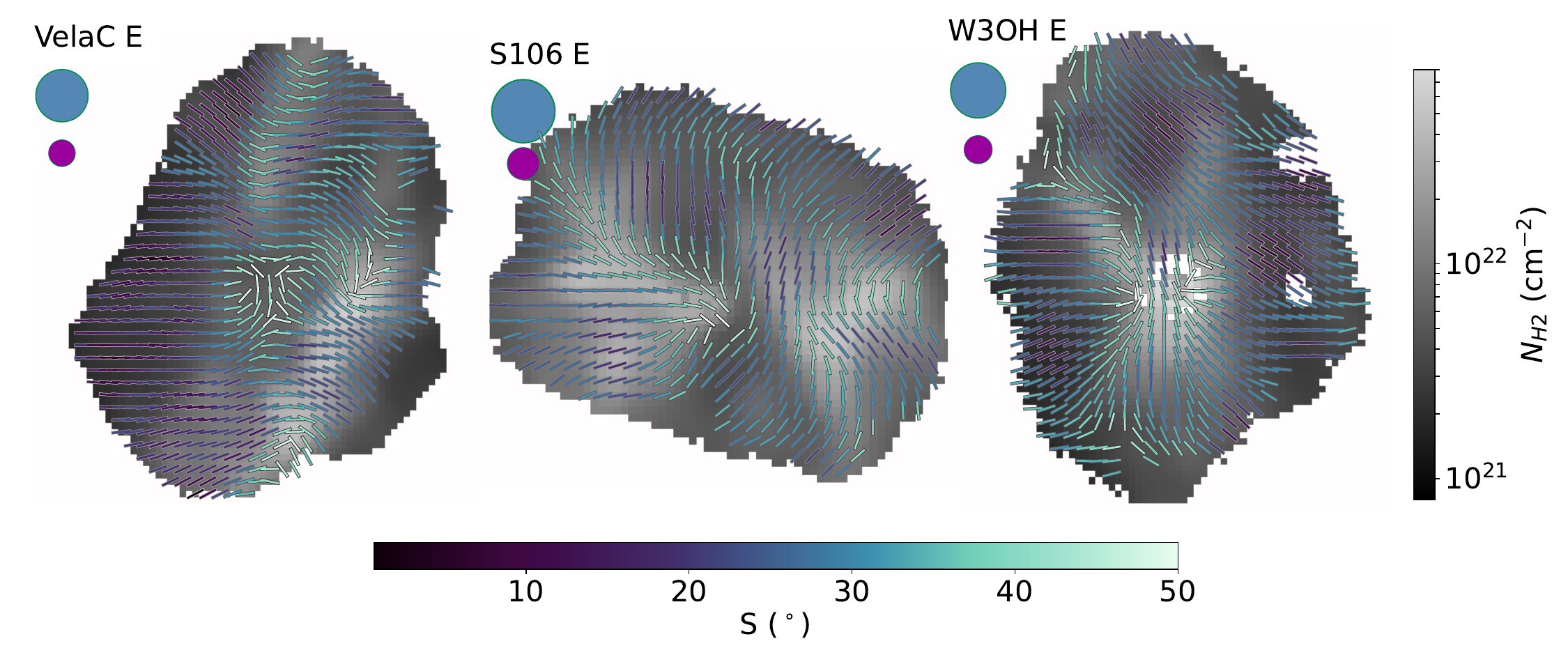}
    \caption{Example maps of the column density with magnetic field pseudovectors overplotted. The vectors are color coded by the angular dispersion and have constant length. The blue circle shows the size of the disk that the dispersion is calculated over and the purple circle shows the $25''$ beam.}
    \label{fig:NH2_dispvectors}
\end{figure*}

In addition to the observed polarization properties, we utilize three derived quantities: the column density, temperature, and polarization angular dispersion. We fit a modified blackbody curve to the Herschel data described in \autoref{subsec:herschel} to estimate the column density, dust temperature, and dust emissivity index. The fit is performed three separate times on Herschel data that is smoothed to the common angular resolution ($25''$) and the common physical resolutions (0.052 pc and 0.32 pc). We follow the method for modified blackbody curve fitting described in \citet{Chuss2019}. 

The intensity is modeled as:
\begin{equation}
    I_{\nu}=(1-e^{-\tau(\nu)})B_{\nu}(T), \label{eq:black}
\end{equation}
where $B_{\nu}(T)$ is the Planck blackbody function at frequency ($\nu$) and temperature ($T$). The optical depth, $\tau(\nu)$, is defined as $\tau \equiv \epsilon(\frac{\nu}{\nu_0})^{\beta}$, where $\epsilon$ is a constant  of proportionality related to the column density along the line of sight and $\beta$ is the dust emissivity index. We use a value of $1000$ GHz for $\nu_0$ and $\epsilon = \kappa_{\nu_0}\mu m_H N_{H_2}$. $\kappa_{\nu_0}$ is a reference dust opacity per unit mass at $\nu_0$, $\mu$ is the mean molecular weight per hydrogen atom, $m_H$ is the atomic mass of hydrogen, and $N(H_2)$ is the gas column density. We use $\kappa_{\nu_0}=0.1 \text{ cm}^2 \text{ g}^{-1}$ and $\mu = 2.8$ following \citet{Sadavoy2013}. 

In star-forming regions, $\beta$ can vary generally between $\sim 1 \text{ and }3$ \citep[e.g.,][]{Hunter1998, Friesen2005, Chuss2019}. These variations can be caused by different dust grain compositions or from fitting a single temperature to a system with multiple temperature components along the line of sight \citep{Draine2006}. Because of this, we do not wish to choose a single constant value of $\beta$ to apply to every sightline in every region. However, there is a fitting degeneracy between $\beta$ and $T$ that makes it hard to fit a modified blackbody, especially for higher temperature objects, with only four or five data points for each pixel \citep{Shetty2009}. We decide to fit for a different $\beta$ value for each region that is then used for every pixel in that region. To do this, we create a footprint from the HAWC+ data that satisfy $I/\sigma_I\ge200$ for any wavelength for a region, and use this as a mask on the Herschel data. The pixels within this mask are then used to find a single $\beta$ value that produces the lowest $\chi^2$ map when used in fitting a map of temperature and column density. For some regions, the fit preferred values of $\beta$ that were below 1, which has been observed in protoplanetary disks likely due to dust grain growth, but is not expected for dust in molecular clouds \citep{Draine2006}. Our data are not sensitive to individual protoplanetary disks, which have radii of 10s to 100s of au \citep{Trapman2023}, so we do not physically expect $\beta$ to be less than 1. We use a flat prior on $\beta>1$, but to discourage fitting $\beta<1$, we use a prior equal to $\beta^2$ for $\beta<1$. A Markov-Chain Monte Carlo (MCMC) method is used to estimate the uncertainty on $\beta$. These values are reported in \autoref{tab:regions}. We then use this constant $\beta$ value and the Herschel intensity maps to fit for maps of the temperature and column density, using a MCMC method to determine their uncertainties. We use the $\beta$ value derived from the $25''$ for the 0.052 pc and 0.32 pc fits. After completing the fits, we calculate the reduced $\chi^2$ statistic for each pixel. For our analyses, we only use pixels where $\chi^2_{reduced}<3$. \autoref{fig:NH2_dispvectors} shows three example plots of our derived column density maps that satisfy the $\chi^2_{reduced}<3$ cut at a resolution of $25''$ with magnetic field pseudovectors overlayed. For some of the Herschel datasets, there are saturation effects at the brightest peaks, resulting in worse fits to the modified blackbody spectrum, {and} those pixels being cut from our analyses. This can be seen in the derived column density map shown in W3OH in \autoref{fig:NH2_dispvectors}, where some of the center pixels were cut. 

\begin{figure*}
    \centering
    \includegraphics[width=\textwidth]{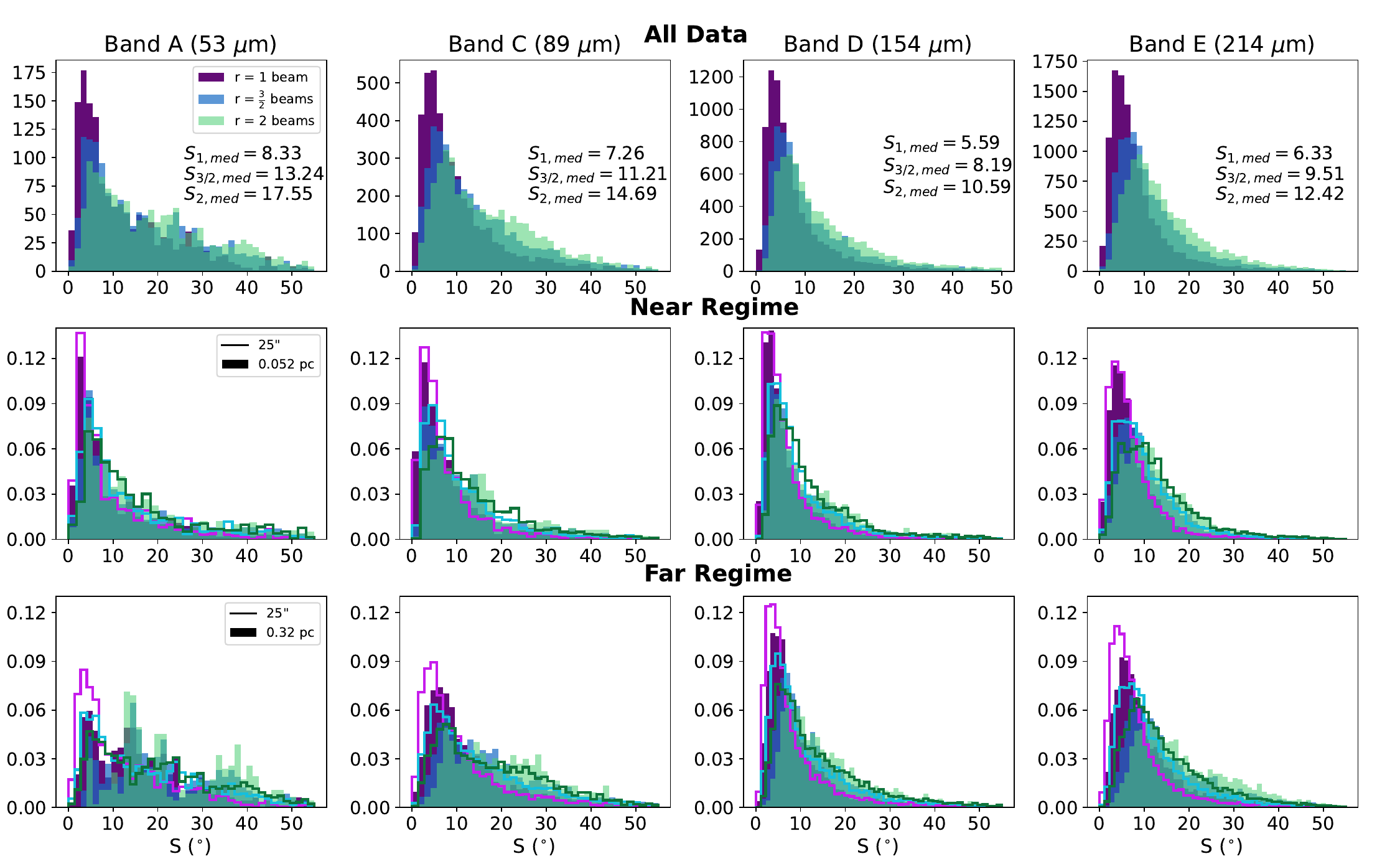}
    \caption{Histograms of the angular dispersion calculated over three different areas with radii of 1, 3/2, and 2 beams. \textit{Top}: All of the data smoothed to $25''$. {The median $S$ value for the three different radii are denoted on each plot} \textit{Middle}: {Normalized histograms of} the near regime data smoothed to $25''$ (solid lines) and smoothed to 0.052 pc (shaded regions). \textit{Bottom}: {Normalized histograms of} the far regime data smoothed to $25''$ (solid lines) and smoothed to 0.32 pc (shaded regions).
    \label{fig:disp_hist}}
\end{figure*}

The per-pixel angular dispersion ($S$) is calculated from the HAWC+ maps as the circular standard deviation of all pixels within a disk around each map pixel. The circular standard deviation for polarization pseudovectors is defined as
\begin{equation}
\label{eq:S}
    S=\frac{1}{2}\sqrt{-2 \ln(\bar R)},
\end{equation}
where $\bar R$ is the resultant length and can be defined as

\begin{equation}
\begin{aligned}
    \bar R = \frac{1}{N}\sqrt{\left( \sum\limits_{n=1}^{N} \sin2\phi_n \right)^2+\left( \sum\limits_{n=1}^{N} \cos2\phi_n \right)^2}.
\end{aligned}
\end{equation}
We calculate $S$ using the linear Stokes parameters where \(\text{sin}2\phi=\frac{U}{\sqrt{Q^2+U^2}}\) and \(\text{cos}2\phi=\frac{Q}{\sqrt{Q^2+U^2}}\) {and $N$ is the number of pixels $S$ is calculated over.}

The uncertainty on $S$ is estimated using a Monte Carlo method, by randomly sampling within the uncertainties on Stokes $Q$ and $U$. $S$ always has a positive value and is biased by noise. $S$ is debiased using its uncertainty
\begin{equation}
    S_{debiased}=\sqrt{S^2-\sigma_S^2},
\end{equation}
but $\sigma_S$ is generally a small quantity (fractions of a degree), so we note that debiasing does not have a noticeable impact on the results in this paper. We use $S_{debiased}$ for all of our analyses and will hereafter refer to it as $S$. We calculate $S$ using radii of 1, $\frac{3}{2}$, and 2 beams. We show the distributions of $S$ in all the data, the {near} regime, and far regime, for each of these radii in \autoref{fig:disp_hist}. Further analysis in \autoref{subsec:pvS} is done using $S$ calculated over a disk of radius of 1 beam. We also calculate $S$ for the smoothed data, where the dispersion is calculated over a constant physical scale that has a diameter of two beams. The pseudovectors in \autoref{fig:NH2_dispvectors} are colored by the value of $S$ calculated over a disk of radius 1 beam ($25''$).  

We note that using the circular standard deviation to calculate $S$ differs slightly from the polarization angle dispersion function used in other works \citep[e.g.,][]{PlanckXII}. We compare our method of calculating $S$ using the circular standard deviation with the polarization angle dispersion function, and the two are highly correlated with an average Pearson $r$ value across all regions of 0.97. The results do not qualitatively change when computed with the conventional polarization angle dispersion function. {See also \citet{Pattle2017} for an alternate method of calculating $S$.}

\section{Results and Discussion}
\label{sec:results}
{In this section, we analyze trends in the polarization fraction as a function of wavelength and resolution (\autoref{subsec:perc_pol}, \autoref{subsec:pol_spec}) and the polarization orientation (\autoref{subsec:bfield}). We also discuss factors that can result in depolarization such as: beam depolarization and magnetic field inclination angle (\autoref{subsec:pvS}) and reduced alignment efficiency in dense regions (\autoref{subsec:pNH2}). In \autoref{subsec:pxS}, we use the $p\times S$ statistic to explore possible evidence of additional depolarization that is not related to the magnetic field geometry.}

\subsection{Percent Polarization}
\label{subsec:perc_pol}
\begin{figure*}
    \centering
    \includegraphics[width=\textwidth]{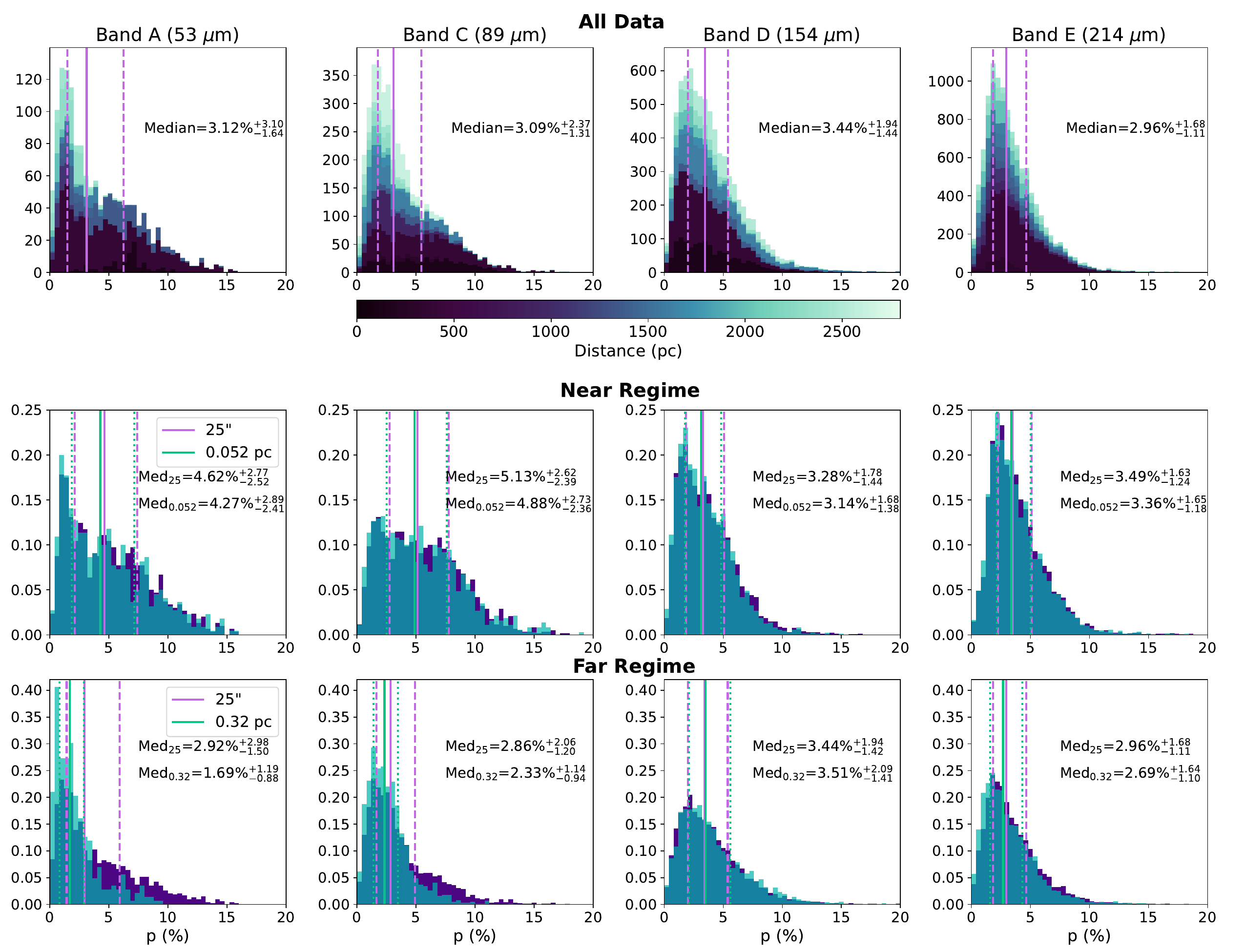}
    \caption{Histograms of the percent polarization. \textit{Top}: All of the data in a stacked histogram smoothed to $25''$. The data are colored based on their distance. \textit{Middle}: {Normalized} histograms of the data in the {near} regime smoothed to $25''$ (purple) and smoothed to 0.052 parsec (cyan). \textit{Bottom}: {Normalized} histograms of the data in the {near} regime smoothed to $25''$ (purple) and smoothed to 0.32 parsec (cyan). The medians with first and third quartiles are denoted in text on the plots. The solid line is the median value and the two segmented lines show the first and third quartiles.}
    \label{fig:pol_hist}
\end{figure*}

\begin{figure}
    \centering
    \includegraphics[width=1\linewidth]{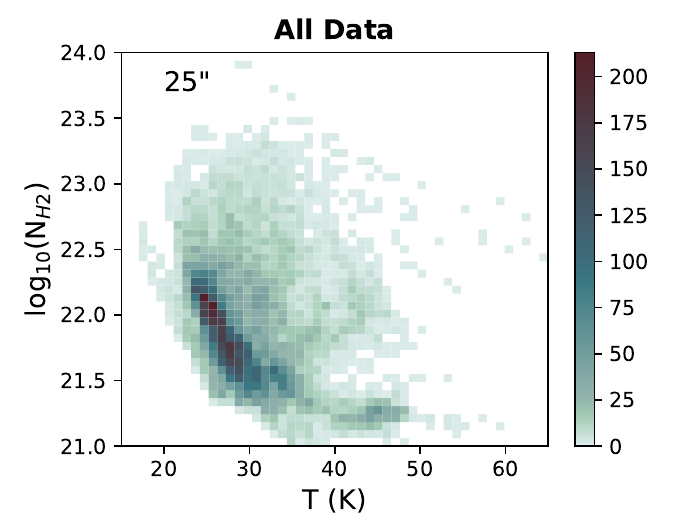}
    \caption{{A 2D histogram of the derived column density versus temperature for all 26 regions in this compilation. The data are at a $25''$ resolution.}}
    \label{fig:TvsNH2}
\end{figure}
With a large population of star-forming regions at various distances and locations in the galaxy, we search for common trends in the plane-of-sky polarization data. We start by investigating the distributions of $p$ in all HAWC+ bands (53-214 \mum) at both common angular and physical resolutions, shown in \autoref{fig:pol_hist}. We report the median, first quartile, and third quartile values on each plot. In the top row, we show a stacked histogram where the data are colored based on the distance of the region they come from. In Bands A and C, we see a generally negative trend between $p$ and distance, where the largest $p$ are in the closest regions. The maximum $p$ detected in each band is $\sim15-20\%$. The highest $p$ values (around $\sim20\%$) are in the Bands D and E data. 

A similar analysis with a sample of 14 star-forming regions was done at wavelengths of 60 \mum~and 100 \mum~from the STOKES polarimeter on the Kuiper Airborne Observatory (KAO) and at 350 \mum~from the Hertz polarimeter on the Caltech Submillimeter Observatory \citep[CSO;][]{Hildebrand_1999}. The resolutions of these observations were $22''$ \citep{Dowell1997}, $35''$ \citep{Platt1991}, and $20''$ \citep{Dowell1998} for 60 \mum, 100 \mum, and 350 \mum~ respectively. This work included multiple nearby regions that we also use in our analysis, but their data also extended to larger distances, including multiple clouds in the Galactic Center. \citet{Hildebrand_1999} observed a distribution of $p$ that was generally in the range $0\%\lesssim p\lesssim10\%$. They also saw the broadest distribution of $p$ at their shortest wavelength of 60 \mum. 

Compared to this previous work, we find a larger range in $p$, as significant portions of our data have values above 10\%. Bands A and C in the {near} regime have the broadest distributions, as quantified by the interquartile range, {with more pronounced high-$p$ tails}  (\autoref{fig:pol_hist}). The far regime also shows a larger interquartile range in Bands A and C in the common angular resolution data, but it is less pronounced than in the {near} regime. {This wavelength dependence of $p$ could imply that grains of different temperatures have different polarization efficiencies, as the shorter wavelength emission is likely dominated by warmer grains. Under the same physical conditions, smaller classical dust grains will have a higher equilibrium temperature than larger grains \citep{Draine2011_textbook}. However, smaller grains tend to exhibit lower polarization efficiency than larger grains \citep{Cho2005}. Since we observe higher $p$  at the shorter wavelengths, we posit that the variations in the $p$ distributions come from the shorter wavelengths preferentially probing a distinct population of warmer dust grains. 
This warmer dust could exhibit a higher fractional polarization due to a higher {grain alignment} efficiency, perhaps enhanced by radiation from embedded sources. Alternatively, the warmer dust could be associated with more ordered magnetic fields, such that the polarized emission from the warmer dust grains experiences less geometric depolarization. 
A similar trend has been seen in external starburst galaxies, where shorter-wavelength far-infrared polarized emission preferentially traces a distinct population of warm dust grains \citep{Lopez-Rodriguez2023}.}  

We also analyze how the resolution of data impacts the distributions of $p$. The distributions of $p$ in the {near} regime do not change greatly when smoothed to a common physical resolution (0.052 pc). This is likely because multiple regions in the {near} regime do not need to be smoothed with large kernels to get to the final resolution. The distributions of $p$ in the far regime change more when smoothed to a common physical resolution (0.32 pc). The most dramatic change is a decrease in the high-$p$ tail of the distributions in Bands A and C. We analyze the distributions of $p$ in each region individually at the common angular resolution. We find that the more nearby regions (smallest physical resolution initially) generally had the highest $p$ at $25''$ resolution in Bands A and C, and then once smoothed to the same physical resolution for the far regime, all the regions had similar $p$ distributions to the farthest dataset. Conversely, the $p$ distribution in Bands D and E in the far regime show no significant change when smoothed to a common physical resolution and also do not show a dependence on distance before smoothing. 

Smoothing the data will generally lower the percent polarization if the new beam size averages over regions of high and low $p$ or varied polarization orientation (beam depolarization). We can see in \autoref{fig:disp_hist} that {Bands A and C have higher median $S$ and more prominent high-$S$ tails}, signifying that there is more variation in the polarization angle orientations. This is consistent with the observation that smoothing to a common physical resolution disproportionately affects the shorter wavelengths. 

Star-forming regions exhibit structure on many scales, from filaments that are generally observed to have widths on the order of $\sim0.1$ pc \citep{Arzoumanian2011, Hennemann2012, Schisano2014, FernandezLopez2014, Panopoulou:2022} to dense cores that often have sizes in the range $\sim0.01-0.1$ pc \citep{Cesaroni2005, Andre2010}. The observed polarization structure of our shorter-wavelength bands could be caused by pockets of warmer dust within our regions, if these are associated with small-scale magnetic field structure. {We examine the $T-N_{H_2}$ distribution at $25''$ resolution for all 26 regions in this compilation (\autoref{fig:TvsNH2}), and confirm that there are sight lines with warm ($T\gtrsim30$ K), dense ($N_{H_2}\gtrsim10^{22}$ cm$^{-2}$) dust that could be in cores and filaments. There is also a tentative trend of increasing $p$ with increasing $T$ {in Bands D and E (\autoref{fig:pvT})}, which is consistent with increased polarization efficiency in warmer grains. However, the column density and temperature derived from a single-component modified blackbody fit are line-of-sight averaged and can be difficult to interpret in sight lines where there are multiple populations of dust. While these fits may be dominated by dense, cold dust, the shorter-wavelength polarization preferentially traces the localized warm component.}

The scale dependence of the polarized dust emission could be further probed by measuring polarization power spectra of these regions \citep[e.g.,][]{PLanckXI, Cordova2024}. Such analyses would be better enabled by high sensitivity observations with larger dynamic range, e.g., as could be achieved with the proposed PRobe Infrared Mission for Astrophysics \citep[PRIMA;][]{Glenn2025}.

\subsection{Polarization Spectra}
\label{subsec:pol_spec}
\begin{figure*}
    \centering
    \includegraphics[width=\linewidth]{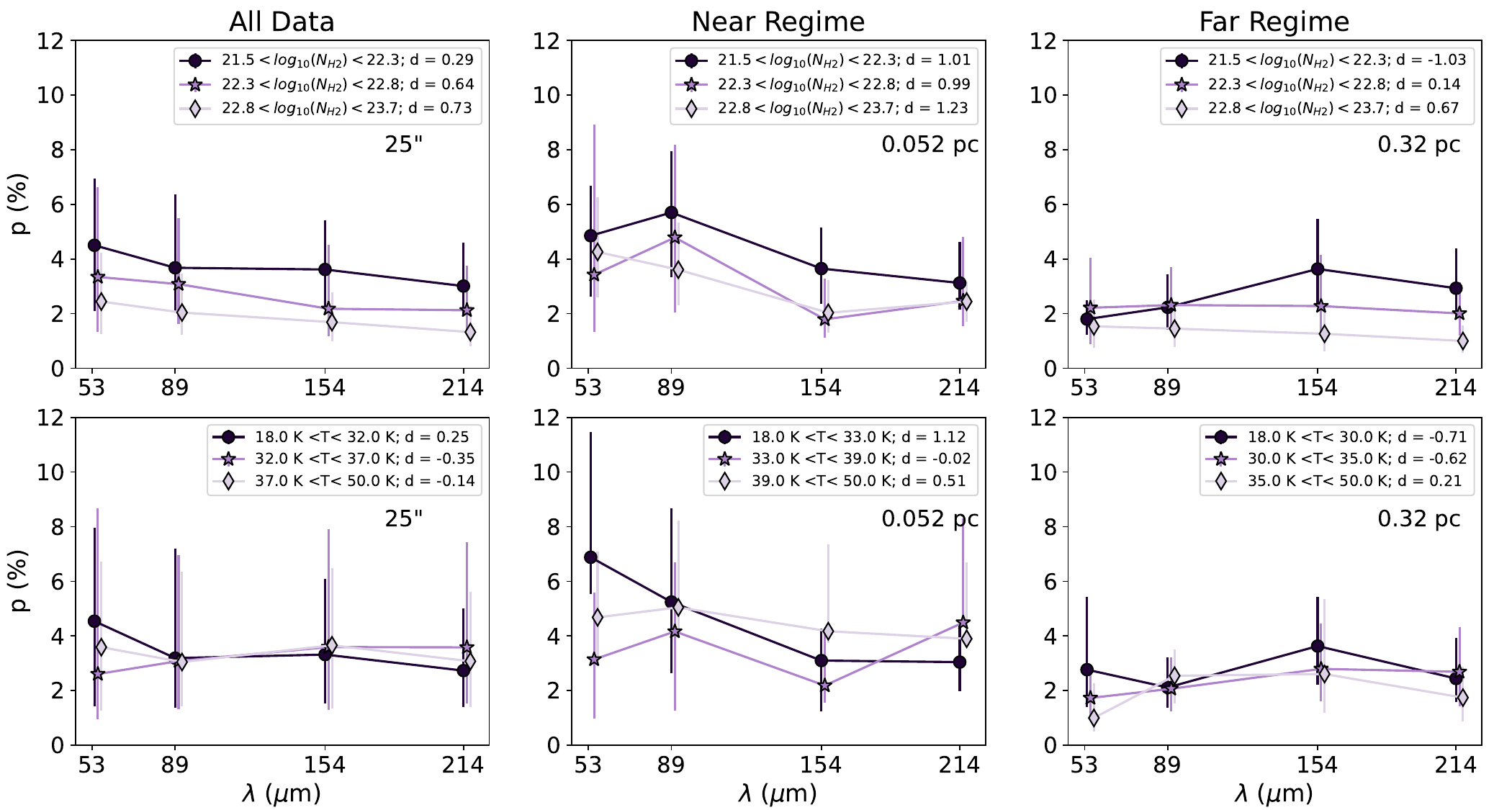 }
    \caption{Polarization spectra created by separating the $p$ data in $N_{H_2}$ (\textit{Top}) and $T$ (\textit{Bottom}) bins using circles, stars, and diamonds for the lowest, middle, and highest bins respectively. The error bars denote the first and third quartiles of the binned data. This is done for all of the data at a resolution of $25''$ (\textit{Left}), the {near} regime at a resolution of 0.052 pc (\textit{Middle}), and the far regime at a resolution of 0.32 pc (\textit{Right}). The bins we use are denoted in the legends. We also denote the Cohen's $d$, calculated as defined in the text, of each spectrum in the legend}
    \label{fig:pol_spectra}
\end{figure*}

\begin{figure*}
    \centering
    \includegraphics[width=\linewidth]{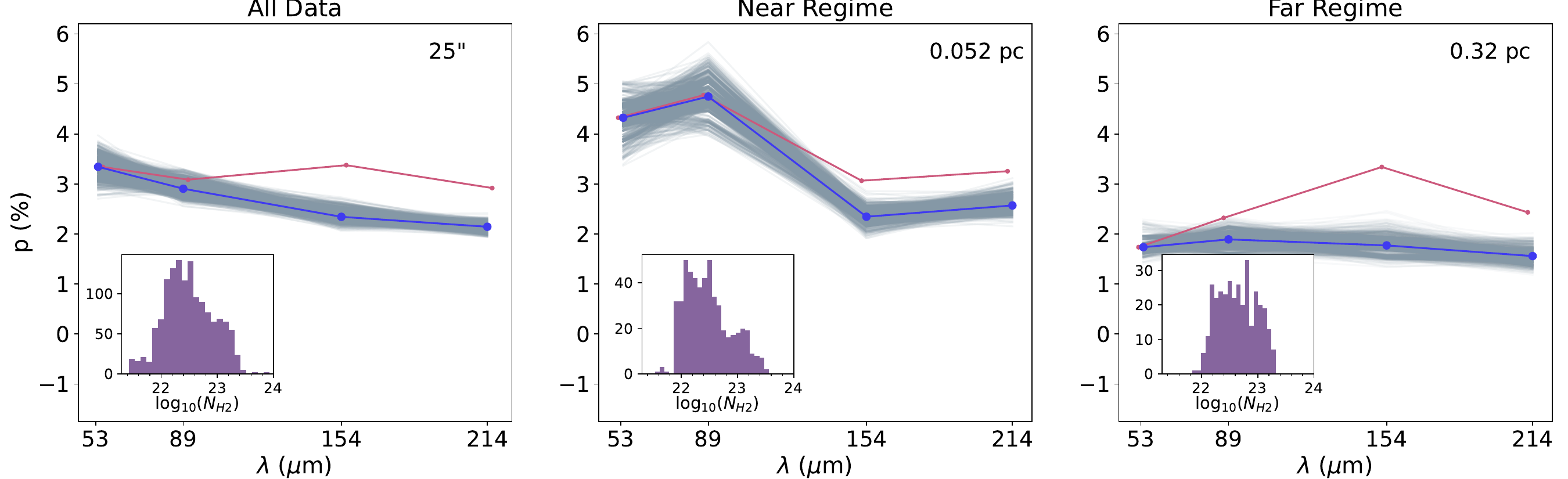}
    \caption{Polarization spectra computed over constant $N_{H_2}$ distributions for all of the data (\textit{left}), the {near} regime smoothed to 0.052 pc (\textit{middle}), and the far regime smoothed to 0.32 pc (\textit{right}). The blue spectra are the median $p$ values of the 1000 samples. The grey lines show the $p$ for each of the 1000 trials. The inset histograms show the distribution of $N_{H_2}$ every band is fixed to. The pink data{points} are the same as the common resolution data medians reported in \autoref{fig:pol_hist}.}
    \label{fig:pol_spectra_NH2}
\end{figure*}

We use our compilation to broadly understand polarization spectra --- the change in $p$ with wavelength --- in Galactic molecular clouds. Using the median values of the $p$ distributions in \autoref{fig:pol_hist}, we observe diverse behaviors across our regimes. In the {near} regime, $p$ is generally higher in Bands A and C, approximating a falling polarization spectrum (decreasing $p$ with increasing wavelength), though the behavior is not strictly monotonic. The behavior of the far regime is distinct, with a median $p$ that is approximately flat or increasing with wavelength. 

Variations in polarization spectra have been seen across different clouds and environments before. \citet{Hildebrand_1999} analyzed polarization spectra in molecular clouds with data from KAO and CSO at 60\,\mum, 100\,\mum, and 350\,\mum. They found that different environments of the cloud may have ``rising" spectra that increase with wavelength, ``flat" spectra that show no change with wavelength, or ``falling" spectra that decrease with wavelength. Using data from KAO, CSO, the Submillimetre Common-User Bolometer Array (SCUBA) polarimeter on the James Clerk Maxwell Telescope (JCMT), \citet{Vaillancourt2002} observed a V-shaped polarization spectrum where $p$ decreases as a function of wavelength from 60 \mum~ to 350 \mum, then increases towards longer wavelengths. The wavelength range in our paper covers the decreasing or falling part of the spectra observed in \citet{Vaillancourt2002}. Simulated polarization spectra have shown that the falling and V-shaped spectra can arise in sightlines where there are multiple dust populations with different temperatures along the line of sight \citep[e.g.,][]{Lee2024, Seifried2025}.

Observing the dependence of the polarization spectrum on cloud properties such as temperature ($T$) and column density ($N_{H_2}$) is a way to probe grain alignment mechanisms. Previous studies disagree on whether polarization spectra are primarily driven by $T$ \citep[e.g.,][]{Michail2021} or $N_{H_2}$ \citep[e.g.,][]{Hildebrand_1999, Santos2019, Lee2024, Cox2025}, with recent work frequently finding falling spectra in higher column density regions. Falling spectra in high-column-density regions may occur because dust grains in cold, dense environments self-shield from the radiation required for efficient grain alignment {\citep[e.g.,][]{Hildebrand_1999}}. 

We look for evidence of a consistent dependence of polarization spectra on $T$ or $N_{H_2}$ by computing spectra that consist of data in certain $T$ and $N_{H_2}$ bins. We repeat this analysis for all of the data at $25''$, the {near} regime at 0.052 pc, and the far regime at 0.32 pc, as shown in \autoref{fig:pol_spectra}. It can be difficult to classify the shape of the spectra with only four wavelengths and varying degrees of spread around each median. We also do not necessarily expect the spectra to vary linearly, so defining the spectra with slopes may not fully capture their shape. Since we see qualitative similarities in the distributions of $p$ between Bands A and C versus Bands D and E, we use the Cohen's $d$ statistic to assess whether the spectra are falling, flat, or rising. 
Cohen's $d$ is calculated as
\begin{equation}
    d= \frac{\text{med}(p_{short}) -\text{med}(p_{long})}{\sigma_{pooled}},
\end{equation}
where $\text{med}(p_{short})$ is the median of the data in Bands A and C and $\text{med}(p_{long})$ is the median of the data in Bands D and E. We define
\begin{equation}
    \sigma_{pooled}= \sqrt{\frac{\text{MAD}(p_{short})^2+\text{MAD}(p_{long})^2}{2}},
\end{equation}
where $\text{MAD}(p_{short})$ and $\text{MAD}(p_{long})$ are the median absolute deviations (MAD) of the short and long wavelengths, respectively. We use the median and MAD to be consistent with previous polarization spectra analyses, and to prevent our results from being biased by outliers. With our sign convention, $d>0$ is a falling spectrum and $d<0$ is a rising spectrum. Following convention, we interpret $|d|\geq0.8$ as a significant difference between the two groups of bands. We denote the calculated $d$ in the legends of \autoref{fig:pol_spectra}.

Overall, we find that most bins in both $T$ and $N_{H_2}$ at all resolutions result in a $|d|<0.8$, meaning we interpret most of our spectra as being consistent with flat. Notably, all $N_{H_2}$ bins in the {near} regime at a resolution of 0.052 pc have significant falling spectra. The lowest $N_{H_2}$ bin in the far regime is also a significant rising spectrum. The only significant $d$ we find in the $T$-binned spectra is in the lowest-temperature bin of the {near} regime. We find that there is more noticeably a trend in the shape and magnitude of the spectra with $N_{H_2}$, but no consistent trend with $T$. Generally, across all regimes, the high $N_{H_2}$ bins have the lowest $p$, and the low $N_{H_2}$ bins have the highest $p$, regardless of wavelength. However, when we bin the data by $T$, we do not see any qualitative trends.

Typical polarization spectra analyses are done with cuts to the data to only include sightlines with coverage in all wavelengths and that do not have large changes in the polarization angle with wavelength for a given sightline \citep[e.g.,][]{Hildebrand_1999, Vaillancourt2002, Michail2021}. {To help mitigate effects from the inclination angle of the magnetic field, these analyses created polarization spectra from the ratio of $p$ in each sightline to the $p$ of the longest wavelength.} However, in our data compilation we do not have uniform wavelength coverage in each region, and only have one wavelength for some regions. {Fortunately, the large sample size in our work has the advantage of effectively marginalizing over the magnetic inclination angle, as supported by our results in \autoref{subsec:pvS}. Furthermore, unlike traditional analyses, our approach does not assume that the inclination angle is constant across different dust populations along a sightline.}

We devise a method to compute the polarization spectrum that is akin to having uniform wavelength coverage by sampling spectra from regions with identical properties in either $T$ or $N_{H_2}$. Since \autoref{fig:pol_spectra} shows a more consistent trend for the shape and magnitude of the spectra with $N_{H_2}$ than $T$, we use $N_{H_2}$ as our control variable and compute spectra where each wavelength has the same $N_{H_2}$ distribution. Because Band A has the fewest sightlines, we use its $N_{H_2}$	distribution as a control. We randomly sample (with replacement) half of the Band A measurements, then draw matching $N_{H_2}$ distributions for the other bands to compute median p values. We do this resampling 1000 times and compute the median of the samples. \autoref{fig:pol_spectra_NH2} shows the spectra created using this method for all of the data, the {near} regime at a common physical resolution, and the far regime at a common physical resolution. We also show faint grey lines for the spectra created for all of the 1000 runs. 

We find that computing spectra from the fixed $N_{H_2}$ samples decreases $p$ in the longer wavelengths compared to calculating the spectra from the full data. This could be because we are fixing the $N_{H_2}$ distribution to the Band A coverage, which tends to target the densest regions of the molecular clouds. High $N_{H_2}$ portions of clouds have been observed to have systematically lower $p$ \citep[e.g.,][]{Fissel2016, PlanckIII}. We see generally consistent spectra across the 1000 samples of our data, as shown by the faint grey lines in \autoref{fig:pol_spectra_NH2}. We find a slight falling spectrum in all of the data at a resolution of $25''$, a more significant falling spectrum in the {near} regime at a resolution of 0.052 pc, and a flat spectrum in the far regime at a resolution of 0.32 pc. 

The transition from a falling spectrum in the {near} regime to the flat spectrum in the far regime may be caused by the physical scales resolved by our common physical resolutions. As noted above, the {near} regime is able to resolve smaller objects like cores and filaments. These are the denser regions of the clouds where we would expect falling spectra if, as argued in \citet{Cox2025}, those spectra are caused by self-shielding in the colder grains. {Our near regime could be resolving cold, high-column density structures that harbor poorly aligned grains, warmer pockets of more efficient grain alignment, or both -- however, the Band A coverage weights our analysis toward the densest, star-forming regions of these clouds, where small-scale warm dust may have an outsized impact on the polarization spectrum.} In this picture, the resolution of the far regime is no longer able to resolve these small structures, and thus the observed polarization spectrum is flatter, and the polarization fraction is lower overall. 

To confirm that the differences we are seeing between the {near} and far regime are not caused by different distributions of $N_{H_2}$ in the different regimes, we repeat the same analysis for all of the data and the {near} regime after fixing to the $N_{H_2}$ distribution of the far regime, and find that the spectra do not change significantly. Even though we see no trend in \autoref{fig:pol_spectra} with $T$, we redo this analysis fixing to the Band A $T$ distribution for all bands and find little difference between the fixed-$T$ spectra and the spectra using all of the data. This confirms that we do not see a strong dependence on $T$ with the polarization spectra of our data.  

\subsection{Magnetic Field Orientation}
\label{subsec:bfield}
\begin{figure*}
    \centering
    \includegraphics[width=\linewidth]{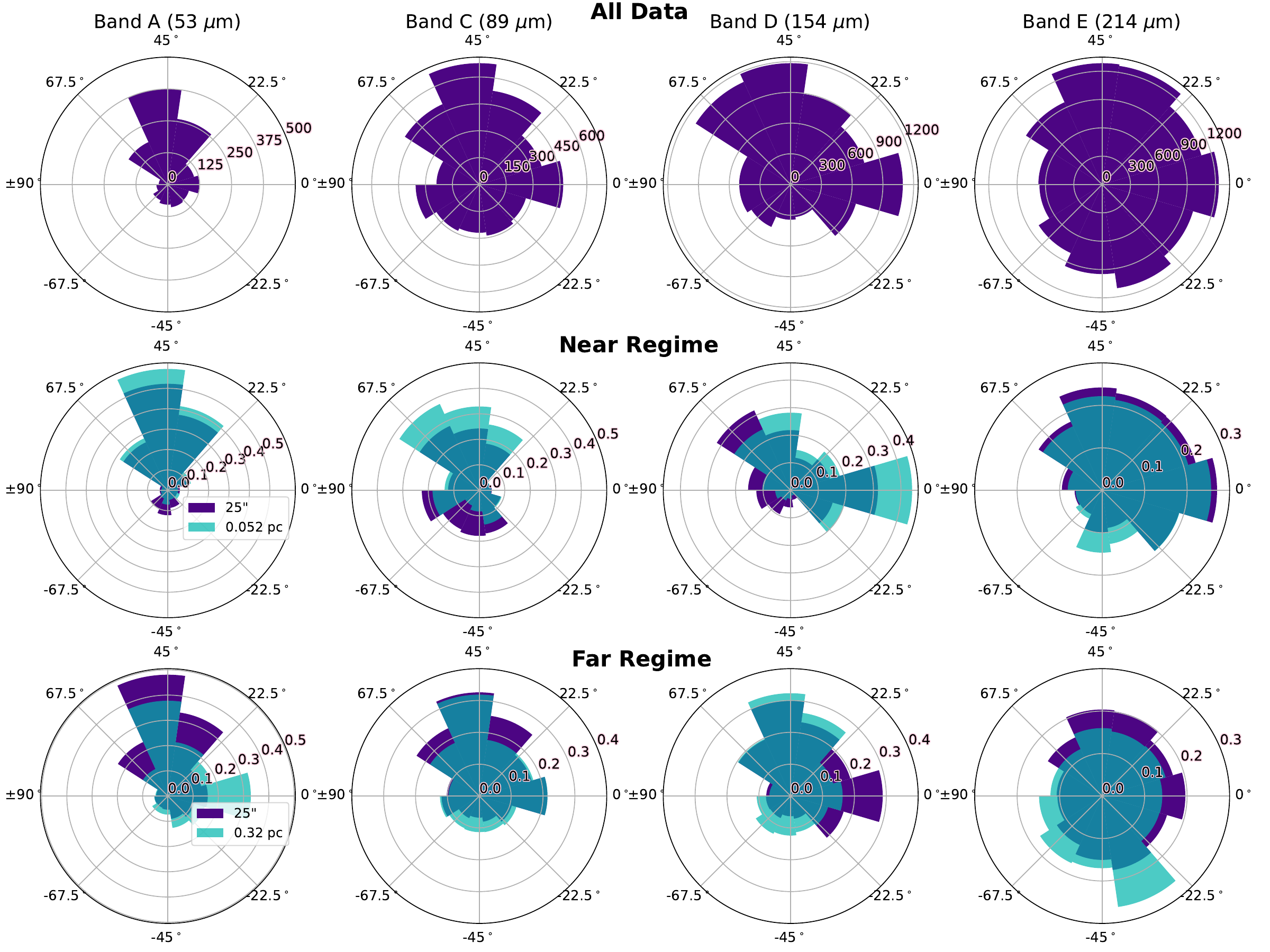}
    \caption{Histograms of the of magnetic field pseudovector orientations ($\phi_B$) in Galactic coordinates. The half-polar polarization data are wrapped onto a full polar plot so that the orientations $+90^\circ$ and $-90^\circ$ are in the same location. An orientation of $0^\circ$ is oriented toward Galactic north. \textit{Top}: All of the data smoothed to $25''$. \textit{Middle}: Density histograms of the data in the {near} regime smoothed to $25''$ (purple) and smoothed to 0.052 parsec (cyan). \textit{Bottom}: Density histograms of the data in the {near} regime smoothed to $25''$ (purple) and smoothed to 0.32 parsec (cyan)}
    \label{fig:angle_hist}
\end{figure*}

We analyze the magnetic field orientation in our regions in the context of larger-scale magnetic fields. In \autoref{fig:fullsky} we show the average magnetic field orientation of each dataset. We calculate these orientations by computing the unweighted average of every sightline in the region that satisfies the signal-to-noise cuts described in \autoref{subsec:HAWC}. A preliminary comparison to the Planck 353 GHz polarization data in each individual region shows no obvious preferred relationship between the Planck-measured magnetic field orientation and the average HAWC+ magnetic field orientation. Our compilation could be used in conjunction with the Planck data to analyze magnetic fields across different spatial scales.

The large-scale Galactic magnetic field in the plane is expected to be parallel with the plane of a spiral galaxy like the Milky Way \citep[e.g.,][]{Han2006}. In agreement with this, \citet{PlanckXIX} found that the average dust polarization orientation in the Galactic plane corresponds to an angle of $\phi=0^\circ$ ($\phi_B=90^\circ$) in Galactic Coordinates. However, the dust polarization orientations in star-forming regions do not necessarily align with the Galactic plane \citep[e.g.,][]{PlanckXIX, Stephens2011}. The magnetic field of the Radcliffe Wave has been observed to be aligned with its structure, resulting in a magnetic field not aligned with the Galactic plane \citep{Panopoulou2025}. The cloud-scale magnetic fields may also become decoupled from the large-scale Galactic magnetic field during the formation of the cloud as it collapses, or from stellar feedback \citep{Pattle2023}. 

The average magnetic field angles of our HAWC+ data reported in \autoref{fig:fullsky} do not show a preferential alignment with $\phi_B=90^\circ$, nor any other orientation. We also examine histograms of the magnetic field orientation (\autoref{fig:angle_hist}). These histograms show no overdensity of measurements in the bins near $\phi_B=\pm90^\circ$. We find no $N_{H_2}$ dependence in the magnetic field orientation, indicating that even the lower $N_{H_2}$ portions of the clouds we cover do not show increased influence from the large-scale Galactic magnetic field.

\subsection{Depolarization and Column Density}
\label{subsec:pNH2}

\begin{figure*}
    \centering
    \includegraphics[width=\textwidth]{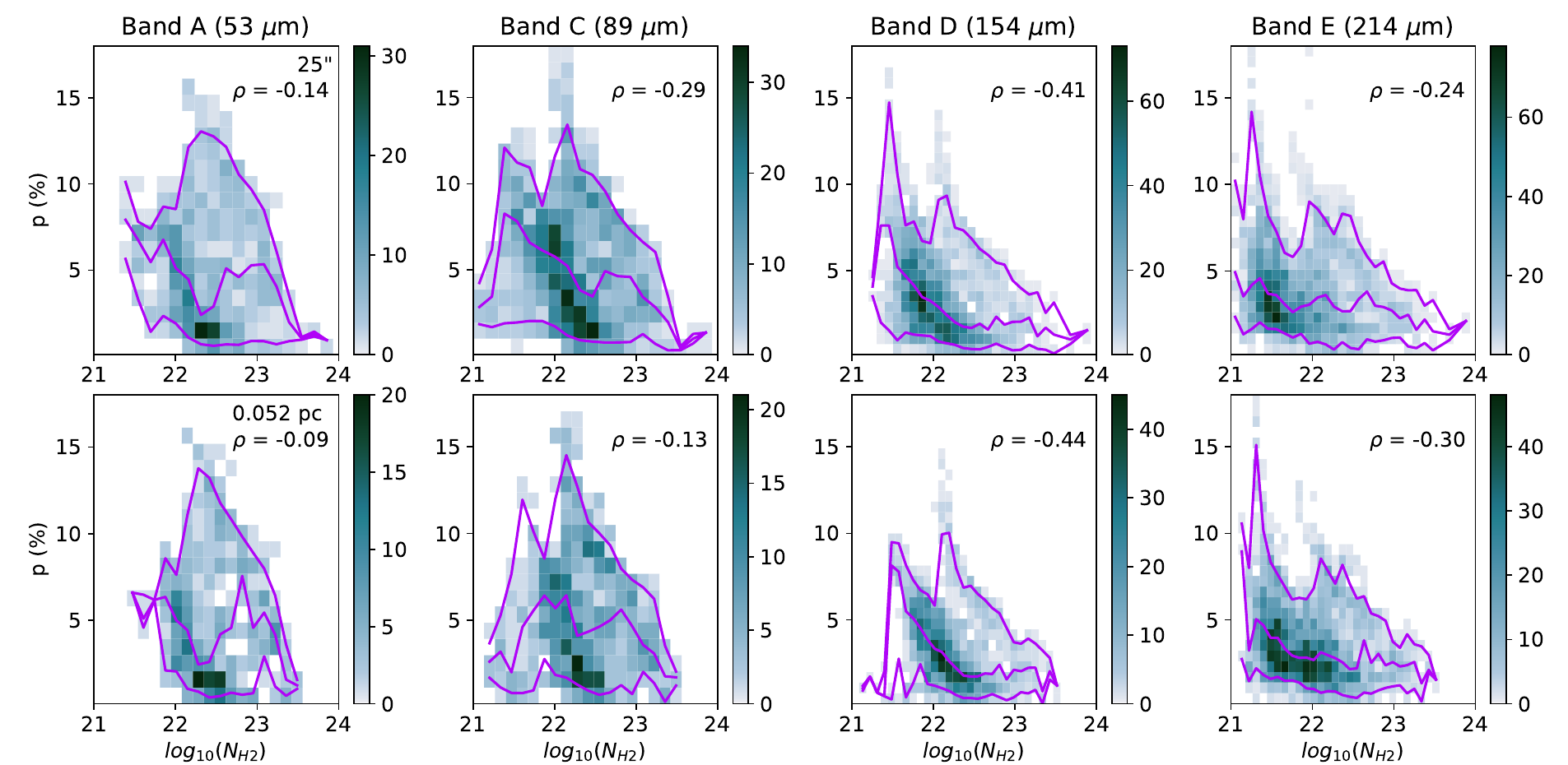}
    \caption{2D histograms of the percent polarization as a function of column density across the four HAWC+ bands for the {near} regime smoothed to $25''$ (\textit{top}) and 0.052 pc \textit{bottom}. The three purple lines from bottom to top are the fifth percentile, median, and ninety-fifth percentile. The Spearman $\rho$ coefficient of the data is reported on each plot to quantigfy the inverse relationship.}
    \label{fig:pvsN_close}
\end{figure*}
\begin{figure*}
    \centering
    \includegraphics[width=\textwidth]{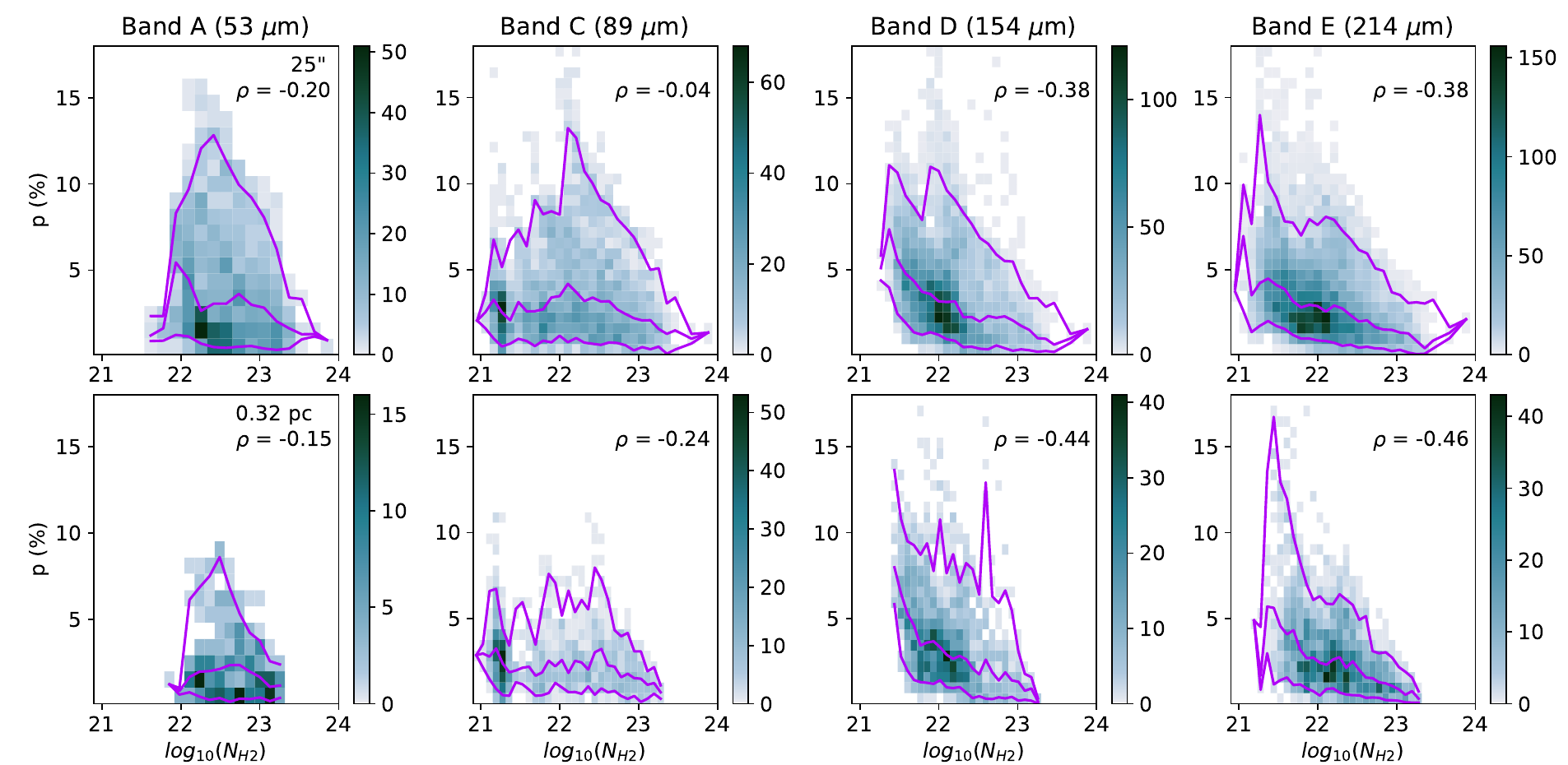}
    \caption{The same as \autoref{fig:pvsN_close} but for the far regime. The bottom row is smoothed to 0.32 pc.}
    \label{fig:pvsN_far}
\end{figure*}
\begin{figure}
    \centering
    \includegraphics[width=\linewidth]{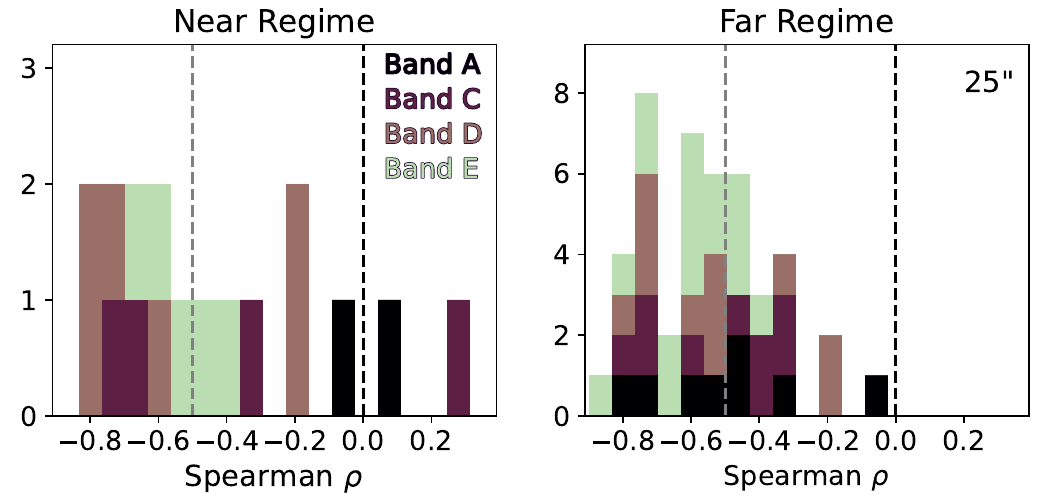}
    \caption{Stacked histograms of the Spearman $\rho$ coefficients between $p$ and $N_{H_2}$ for the individual datasets separated into the {near} (\textit{left}) and far (\textit{right}) regimes at the common angular resolution. The data in each band are stacked and are colored blue, green, orange, and red for Bands A, C, D, and E respectively. The grey dashed line is at $\rho=-0.5$ and the black dashed line is at $\rho=0$.}
    \label{fig:NH2_rho}
\end{figure}
We analyze the relationship between $p$ and $N_{H_2}$ in the {near} (\autoref{fig:pvsN_close}) and far (\autoref{fig:pvsN_far}) regimes using 2-dimensional histograms. Before plotting, we Nyquist sample the data to avoid the appearance of trends from oversampling the beam. We show the fifth percentile, median, and ninety-fifth percentiles as solid magenta lines and report the Spearman $\rho$ value for the data in each plot in \autoref{fig:pvsN_close} and \autoref{fig:pvsN_far}. In both the {near} and far regimes, we broadly see a negative correlation denoted by the calculated Spearman $\rho$ in the range $-0.04$ to $-0.46$, but we do not see any very strong negative relationships of Spearman $\rho\leq-0.5$. In the {near} regime (\autoref{fig:pvsN_close}), the most significant relationship we see occurs in Band D, with $\rho\sim-0.4$. In the far regime (\autoref{fig:pvsN_far}), we see the most negative relationships in Bands D and E at both resolutions, also with $\rho\sim-0.4$.  

Bands A and C generally have the least significant $\rho$ values {for the $p-N_{H_2}$ distributions}, except for Band C in the {near} regime at the common angular resolution (\autoref{fig:pvsN_close}). The ninety-fifth percentiles at higher $N_{H_2}$ ($\sim10^{23}$ cm$^{-2}$), especially in the common angular resolution data, are also generally higher in Bands A and C than in Bands D and E. This implies that warm dust grains have less depolarization in high $N_{H_2}$ regions.

The dust polarization fraction and $N_{H_2}$ have been observed to have an inverse relationship in molecular clouds at longer wavelengths \citep[e.g., 250 -- 850 \mum;][]{PlanckXII,Fissel2016}. We note that our longest wavelengths, especially in the far regime, have the strongest negative correlations, implying that there may be a wavelength dependence to this correlation strength. {Within RAT theory}, a decrease in polarization efficiency in high $N_{H_2}$ regions could be caused by the shielding of radiation preventing the grains from being efficiently radiatively torqued \citep{Hoang2021}, {and/}or from collisions between the dust and gas in high density regions \citep{Purcell1979}. This depolarization could instead be geometric, if the magnetic field is more tangled at high column density \citep{Falceta-Gonclaves2008}. {We consider whether the most significant $p-N_{H_2}$ anticorrelations also display strong $S-N_{H_2}$ correlations, as might be expected if the $p-N_{H_2}$ relationship is driven by geometric depolarization in high $N_{H_2}$ regions. However, we see no strong correlations between $S$ and $N_{H_2}$.} {In \autoref{subsec:pxS} we explore how the quantity $p\times S$ varies with $N_{H_2}$ to help remove the effect of magnetic field tangling.} 

\citet{PlanckXII} found that individual datasets showed different relationships between $p$ and $N_{H_2}$, so we analyze the $p-N_{H_2}$ correlation in each individual dataset. \autoref{fig:NH2_rho} shows a histogram of the Spearman $\rho$ coefficients of the individual datasets smoothed to the common angular resolution. We find that for the common angular resolution, $53\%$ of the {near} regime datasets and $64\%$ of the far regime regions have significant negative correlations of $\rho<-0.5$. For the common physical resolution we find that $44\%$ of datasets in the {near} regime and $64\%$ of datasets in the far regime have significant negative correlations. To estimate uncertainty on $\rho$, we resample the data with replacement 500 times to estimate the $95\%$ confidence intervals. We find that the median widths of these confidence intervals are 0.13 ({near} common angular), 0.12 (far common angular), 0.18 ({near} common physical), and 0.27 (far common physical). Using the higher and lower limits of the confidence interval, we find that $41\%-59\%$ ({near} common angular), $45\%-80\%$ (far common angular), $28\%-56\%$ ({near} common physical), and $40\%-89\%$ (far common physical) satisfy $\rho<-0.5$.

\autoref{fig:NH2_rho} shows that over half of the regions in both regimes have $\rho<-0.5$. Even conservatively using the upper limits of the confidence intervals, we would conclude that a significant fraction of the regions have this strong negative correlation. Just as in the aggregate data, Bands D and E show the strongest $p-N_{H_2}$ anticorrelations, and Bands A and C show the weakest.

We consider whether the datasets with significant $p-N_{H_2}$ anticorrelations also display strong $S-N_{H_2}$ correlations that would drive geometric depolarization in high $N_{H_2}$ regions. This is not clearly supported by the data, however; we find that at the common angular resolution, only $11\%$ ({near} regime) and $36\%$ (far regime) of the datasets with significant $p-N_{H_2}$ anticorrelations also display significant $S-N_{H_2}$ correlations. The mechanisms that drive depolarization may therefore be complex, or variable across regions and/or wavelengths. 

\subsection{Depolarization and Angular Dispersion}
\label{subsec:pvS}
\begin{figure*}
    \centering
    \includegraphics[width=\textwidth]{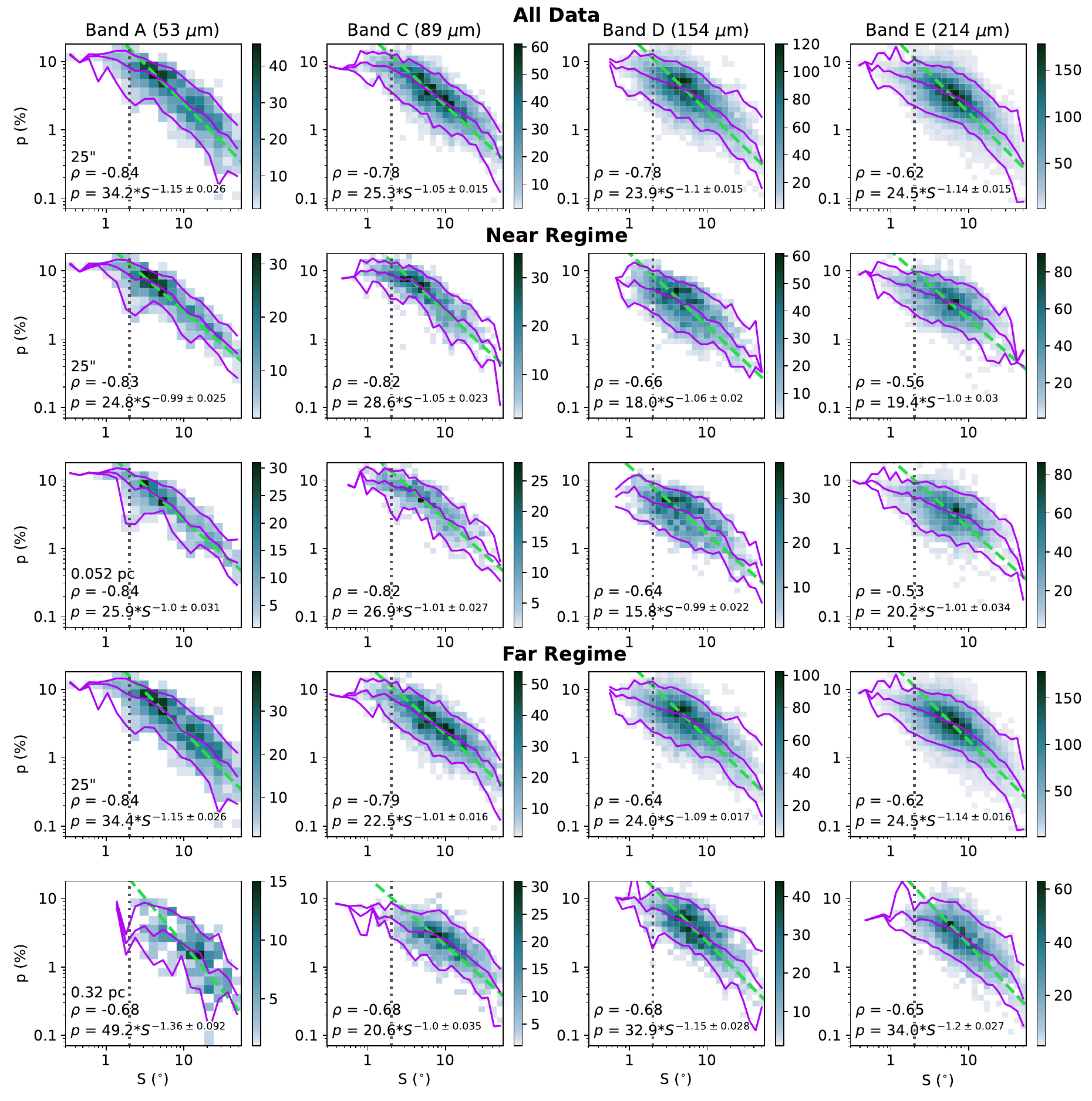}
    \caption{2-D histograms of $p$ as a function of $S$ separated by HAWC+ band. The three purple lines from bottom to top are the fifth percentile, median, and ninety-fifth percentile. The dashed green line shows the power law {derived from the principle axis as} described in \autoref{subsec:pvS}. The {resulting} equation and Spearman $\rho$ are denoted on each subplot. {The  vertical grey dotted lines show the location of the $S<2^\circ$ cut we use when to determine the principal axis of the distribution.}}
    \label{fig:pvsS}
\end{figure*}

We investigate the distributions of $S$. In \autoref{fig:disp_hist}, we show the distributions of $S$ when calculated using radii of 1, $\frac{3}{2}$, and 2 beams for data smoothed to the common angular and physical resolutions. The medians of the distributions of $S${, denoted for the common angular resolution data in the top row of \autoref{fig:disp_hist}}, all increase as the radius {over which} $S$ is calculated over increases. This implies that there are larger magnetic field fluctuations on scales larger than one beam. This trend is seen in both the {near} and far regime, meaning $S$ seems to increase from scales $\sim$ 0.1 pc to $\sim$ 1 pc. The common angular resolution data in the far regime has higher dispersion than the {near} regime, which is consistent with the general decrease in $p$ observed when the far regime data were smoothed to the common physical resolution. 

{There is a trend in the median value of $S$ with wavelength in \autoref{fig:disp_hist}. At the three radii analyzed in all 26 regions at a resolution of $25''$, Bands A and C have the highest median $S$ and Band D has the lowest $S$. A similar trend persists for all radii and resolutions in both the near and far regimes, with the highest and lowest medians in Band A and Band D, respectively. Although we observe a wavelength dependence in $S$, we do not see any significant trends between $S$ and $T$ at any wavelength. This is not inconsistent because the temperature map is line-of-sight density averaged across the cold and warm components, whereas the shorter polarization wavelengths predominantly trace the warm grains.} 

In molecular clouds, $p$ and $S$ have been observed to have an inverse power law relationship {of the form
\begin{equation}
    p = C S^\alpha
\end{equation}}
{with consistent indices ($\alpha$)} across different regions \citep[e.g.,][]{PlanckXII,Chuss2019,Fissel2016}. Power law fits in \citet{Chuss2019} and \citet{Fissel2016} find $\alpha$ between -0.67 and -0.9. \citet{PlanckXII} derived a power law index for an inverse relationship entirely caused by beam depolarization of $\alpha=-1$ \citep[][Appendix F.6]{PlanckXII}. We show 2-D histograms of $p$ as a function of $S$ in \autoref{fig:pvsS} with the fifth percentile, median, and ninety-fifth percentiles as solid magenta lines. {Since the results of linear regression can depend on which variable is chosen as the independent variable, we determine the power law coefficients by following the method employed in \citet{King2018} and calculating the principal axis of the $\text{log}_{10}p - \text{log}_{10}S$ distributions. $\alpha$ is calculated as the ratio of the y component to the x component of the principal eigenvector of the covariance matrix. The uncertainty on $\alpha$ is determined by bootstrapping. We resample with replacement 5000 times and report the 68\% confidence interval as the 1$\sigma$ uncertainties.} Because the power-law relationship breaks down and $p$ flattens at low S, we exclude data where $S<2^\circ$. The results of the fit are reported in \autoref{fig:pvsS}. 

We find significant inverse relationships with our data at all wavelengths and resolutions. We expect an inverse relationship between $p$ and $S$ due to this beam depolarization effect. In \autoref{fig:pvsS}, we show our distribution of $p$ versus $S$ and see {that $\alpha$ ranges from -0.99 to -1.36.} {Fitting for $\alpha$ using $\chi^2$ minimization results in a range from -0.51 to -0.85, consistent with other measurements using linear regression \citep{Fissel2016, Chuss2019}.} The $p-S$ correlation is uniformly negative, and more consistent across regions than the $p-N_{H_2}$ correlation. {We see no qualitative change in $\alpha$ when we change the radius that $S$ is calculated over. Instead we see an increase in the coefficient $C$ as the radius is increased.}

At the common angular resolution, Bands A and C have the lowest indices and strongest negative correlation (Spearman $\rho \sim -0.8$). This contrasts with the observation that the shorter wavelengths had the least significant correlation between $p$ and $N_{H_2}$ (\autoref{subsec:pNH2}). Overall, $p$ shows a stronger negative correlation with $S$ than $N_{H_2}$, meaning a significant amount of the depolarization is likely from beam depolarization. When smoothed to the common physical resolution, there is not much change in the indices or Spearman $\rho$ values in the {near} regime. However, in the far regime the scatter in Bands A and C increases and $\rho$ becomes less significant. {For Bands A, D, and E, the $\alpha$ values are lower than -1 in the far regime at both $25''$ and 0.032 pc resolutions. One reason $\alpha$ may be less than -1 is that there is more dispersion within the beam than within the radius $S$ is calculated over i.e, there is additional depolarization that $S$ is not sensitive to. This supports our argument that the magnetic field varies on smaller scales than we can resolve in the far regime data.} We note that the inclination angle between the mean magnetic field and the line of sight also affects the $p-S$ correlation \citep[e.g.,][]{King2018}, but the uniformity of $\alpha$ across our sample {and its similarity to the value derived in \citet[][Appendix F.6]{PlanckXII} for beam depolarization alone} suggests that inclination effects are not the dominant driver of this correlation slope, as we {expect a range of mean magnetic field inclination angles.}

\subsection{$p \times S $ Relationships}
\label{subsec:pxS}
\begin{figure*}
    \centering
    \includegraphics[width=\textwidth]{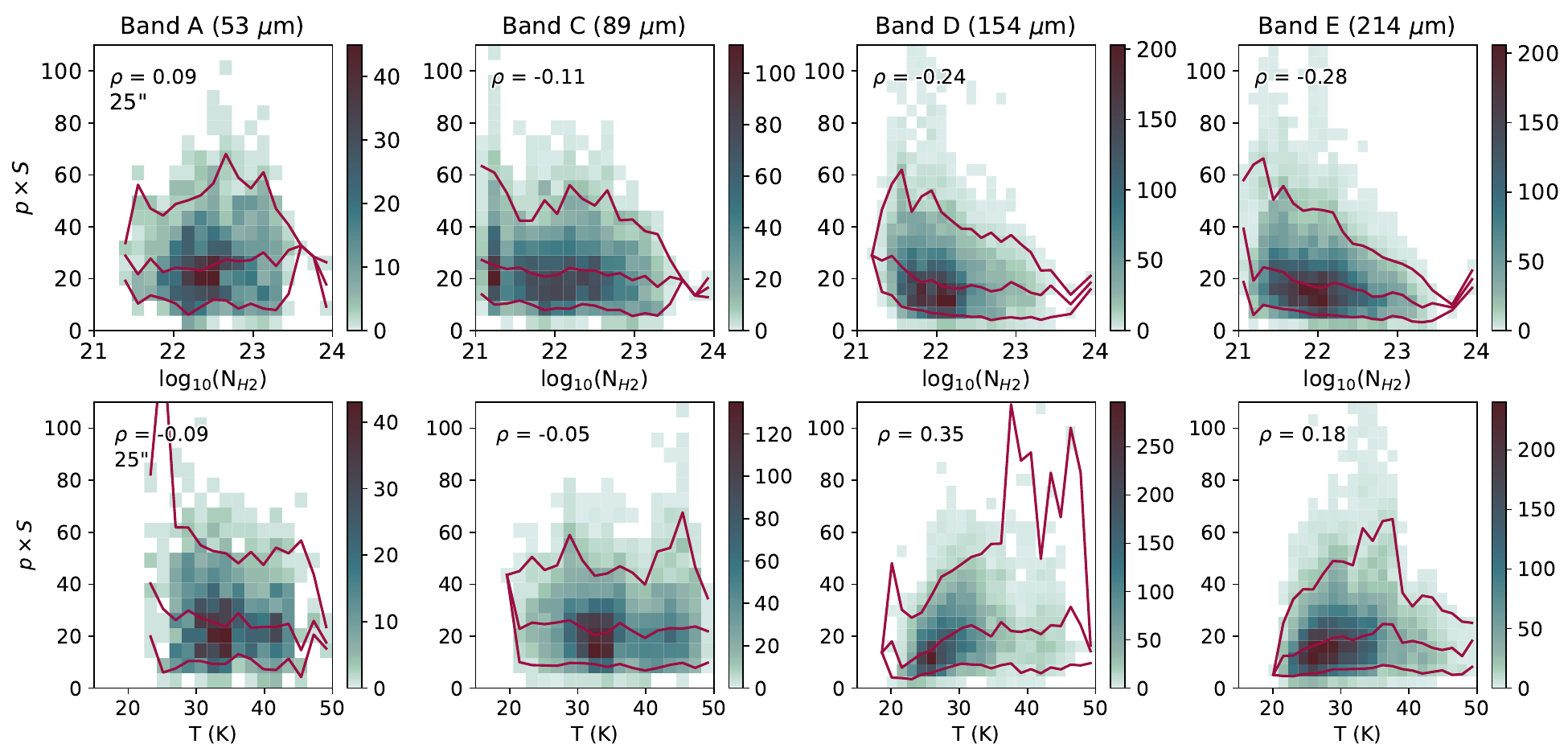}
    \caption{{2D histograms of the $p\times S$ statistic as a function of $N_{H_2}$ (\textit{Top}) and $T$ (\textit{Bottom}). {The three red lines from bottom to top are the fifth percentile, median, and ninety-fifth percentile.} These plots include data from all 26 regions at a resolution of $25''$. The calculated Spearman $\rho$ value is denoted on each plot.}}
    \label{fig:pxS}
\end{figure*}

{In an effort to study the underlying variations in $p$ that are not due to beam depolarization, past works \citep[e.g.,][]{PlanckXII} have analyzed how the value $p\times S$ varies as a function of $T$ and $N_{H_2}$. Since we recover values for $\alpha\lesssim-1$ in \autoref{subsec:pvS} and determine that beam depolarization is the primary source of depolarization correlated with $S$, we create 2D histogram plots of $p\times S$ as a function of $T$ and $N_{H_2}$ in \autoref{fig:pxS} and denote the Spearman $\rho$ for each distribution. We show all the data in each HAWC+ band at a resolution of $25''$. There are not qualitative differences once the data are smoothed to the common physical resolutions.}

{\citet{PlanckXII} found that the quantity $p\times S$ was mostly constant in the column density range $10^{20}<N_{H_2}<2\times 10^{22}$ cm$^{-2}$. We probe a similar regime at higher angular resolution: only 21.7\% of our sight lines have a column density higher than this range. We do not see a strong correlation between $p\times S$ and $N_{H_2}$ in Bands A and C ($|\rho|\lesssim0.11$), while Bands D and E exhibit a weak negative correlation ($\rho$ values of -0.24 and -0.28, respectively). These relationships are less significant than the $p-N_{H_2}$ relationships in \autoref{fig:pvsN_close} and \autoref{fig:pvsN_far}. This is likely because the depolarization in some regions was able to be explained by significant positive correlations between $S$ and $N_{H_2}$ as detailed in \autoref{subsec:pNH2}. The $p\times S-T$ relationship shows a similar trend: Bands A and C are uncorrelated with $|\rho|\lesssim0.1$ and the more significant relationships are in Bands D and E with $\rho$ values of 0.35 and 0.18, respectively. The tentative negative $p\times S-N_{H_2}$ relationship for Bands D and E may be indicative of a loss of grain alignment in denser regions, and the tentative positive $p\times S-T$ relationship may be evidence of increased grain alignment in warmer regions. However, as argued in \citet{PlanckXII}, the low significance of these relationships points towards magnetic field tangling being the predominant source of depolarization rather than loss of grain alignment. The divergence of Bands A and C from these trends may indicate that the warmer grains traced by the shorter wavelengths are less sensitive to changes in temperature and column density. In contrast, the longer wavelengths are a combination of emission from both the well aligned warm and possibly less aligned cold grains, creating the tentative trends with environmental factors.}

\section{Conclusions}
\label{sec:conclusion}
We present archival polarization data from the HAWC+ instrument on SOFIA toward 26 nearby star-forming regions. Far-infrared data from the PACS and SPIRE instruments on the Herschel Space Observatory are used to derive maps of column density ($N_{H_2}$) and temperature ($T$). We analyze trends in the polarization data as a function of wavelength, and the relationship between the dust polarization fraction and $N_{H_2}$, $T$, and polarization angle dispersion ($S$). By analyzing these trends at both a common angular resolution ($25''$) and common physical resolutions (0.052 pc and 0.32 pc), we come to the following conclusions:
\begin{enumerate}
    \item We find that the two shortest wavelengths of HAWC+ (53 \mum~and 89 \mum) exhibit qualitatively distinct properties from the two longer wavelengths. 
    At the common angular resolution, we find that $p$ is significantly affected by the cloud distance, with the $p$ distribution skewed toward higher values for more nearby regions. The shorter wavelengths have wider distributions of $p$ than the longer wavelengths, and smoothing to the common physical resolution decreases $p$ more in the shorter wavelengths. We interpret this as evidence that the shorter wavelengths are sensitive to a distinct dust population: likely warmer grains that may not be spatially co-located with cooler dust populations.  
    
    \item When binned by $N_{H_2}$ and $T$, the shape and magnitude of the polarization spectra calculated from our compilation show a more consistent dependence on $N_{H_2}$ than $T$. We create polarization spectra with identical distributions of $N_{H_2}$ to control for polarization effects that are correlated with $N_{H_2}$. We find a falling spectrum (decreasing with increasing wavelength) when analyzing the data at 0.052 pc, but a flat spectrum at 0.32 pc. The falling spectrum in the {near} regime can be from environmental factors such as shielding of colder grains, or it can be from beam depolarization in these shortest wavelengths. 
    
    \item We find no preferred alignment between the mean plane-of-sky magnetic field in our regions and the Galactic plane. Assuming that the large-scale Galactic magnetic field is parallel to the Galactic plane, our findings imply that the magnetic fields in the star-forming regions we analyze are decoupled from the large-scale Galactic magnetic field.
    
    \item When looking at all the data in each distance regime combined, we consistently see a negative trend between $p$ and $N_{H_2}$, but we do not see any highly significant anticorrelations. However, we do find some significant correlations (Spearman $\rho\leq-0.5$) between these parameters in individual data sets. The individual $p-N_{H_2}$ correlations are not as uniform across regions as the $p-S$ correlations. There can be variations between clouds in their geometry, magnetic field structure, or dust properties that may affect the relationship between depolarization and column density, making it important to explore the physics of these clouds independently.
    
    \item In contrast, the negative relationship between $p$ and $S$ is evident across all wavelengths and resolutions and is a quite consistent trend, with all the data generally falling along one curve{, with power law indices in the range -0.99 to -1.36.} This likely means that beam depolarization is generally a stronger source of depolarization than effects related to $N_{H_2}$ {or a change in grain alignment}. {This is supported by the lack of significant trends between $p\times S$ and $T$ or $N_{H_2}$.}
\end{enumerate}{}
 
Our compilation enables statistical investigations of dust polarization properties in star-forming regions across far-infrared wavelengths. Our results illustrate the importance of observing star-forming regions at both shorter wavelengths ($\lesssim$100 \mum) and longer wavelengths ($\gtrsim$ 100 \mum)  and resolving sub-parsec structures in the dust polarization in order to interpret the observations. Future far-infrared instruments such as PRIMA \citep{Glenn2025}, with its instrument PRIMAger \citep{Ciesla2025}, will be an excellent tool for measuring dust polarization in molecular clouds, with similar resolution and higher sensitivity than HAWC+. PRIMA can be used to analyze a broader range of Galactic environments and the connection between these dense, small scale portions of clouds and larger, more diffuse cloud environments.

\begin{acknowledgments}
We thank Laura Fissel, Brandon Hensley, and Simon Coude for helpful discussions that improved this paper.

Based on observations made with the NASA/DLR Stratospheric Observatory for Infrared Astronomy (SOFIA). SOFIA is jointly operated by the Universities Space Research Association, Inc. (USRA), under NASA contract NNA17BF53C , and the Deutsches SOFIA Institut (DSI) under DLR contract 50 OK 2002 to the University of Stuttgart.

This work was supported by NASA award 80NSSC23K0972. S.E.C. additionally acknowledges support from an Alfred P. Sloan Research Fellowship and NSF CAREER award No. AST-2441452.

E.L.-R. thanks to the support of the NASA Astrophysics Decadal Survey Precursor Science (ADSPS) Program (NNH22ZDA001N-ADSPS) with ID 22-ADSPS22-0009 and agreement number 80NSSC23K1585.

Some of the computing for this project was performed on the Sherlock cluster. We would like to thank Stanford University and the Stanford Research Computing Center for providing computational resources and support that contributed to these research results.

The authors would also like
to thank the anonymous referee for their helpful insight on this manuscript.
\end{acknowledgments}

\facilities{SOFIA, Herschel}

\software{Astropy \citep{astropy:2013, astropy:2018, astropy:2022}, emcee \citep{foreman_mackey2013}, Matplotlib \citep{Hunter:2007}, SciPy \citep{2020SciPy-NMeth}}

\appendix

\renewcommand{\thesection}{A.\arabic{section}}
\renewcommand{\thefigure}{A.\arabic{figure}}
\renewcommand{\thetable}{A.\arabic{table}}
\setcounter{table}{0}
\setcounter{figure}{0}
\section{Supplementary Materials}
\begin{figure*}
    \centering
    \includegraphics[width=1\linewidth]{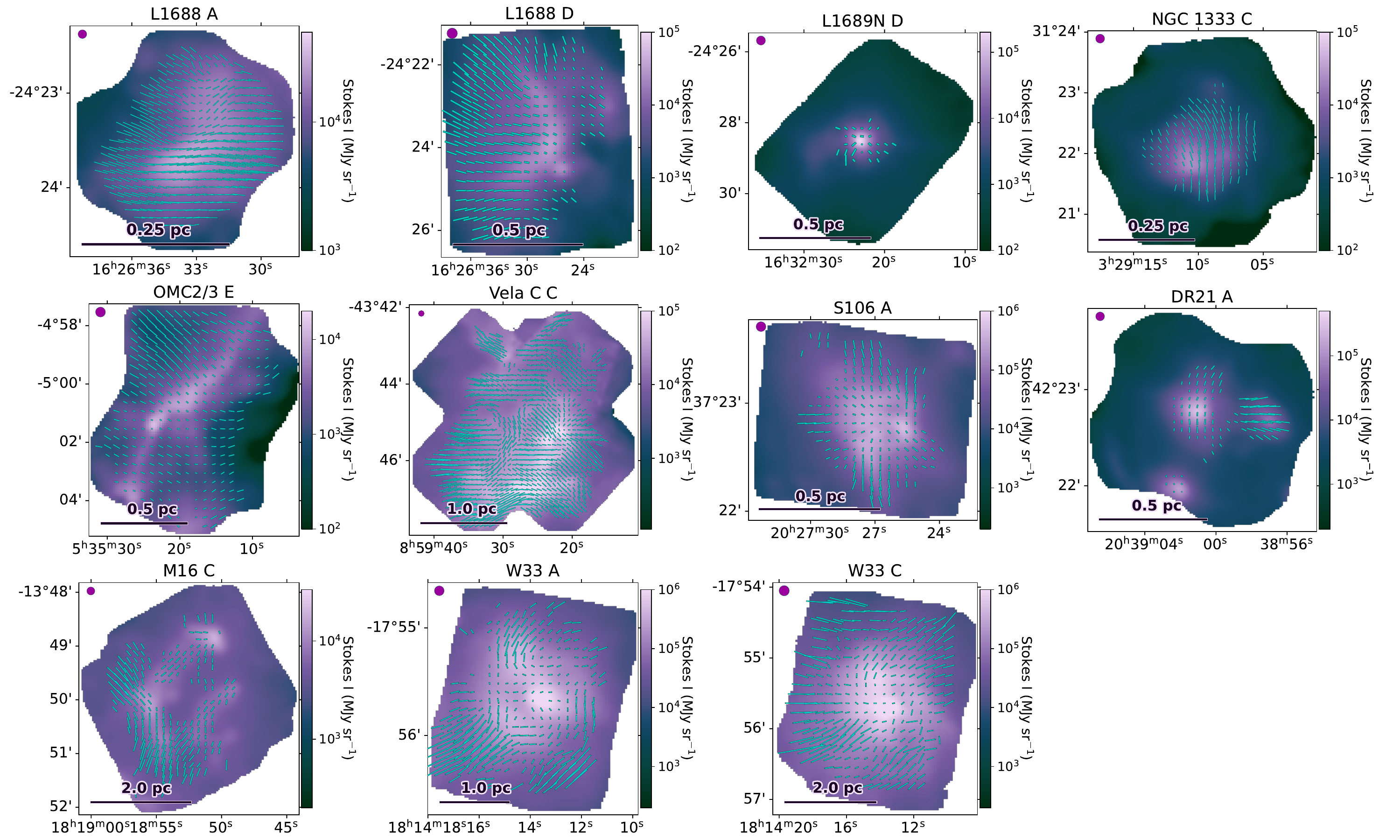}
    \caption{Remaining HAWC+ dataset that are not shown in \autoref{fig:fam_port} or \autoref{fig:four_band}.}
    \label{fig:remaining_data}
\end{figure*}

\begin{figure*}
    \centering
    \includegraphics[width=1\linewidth]{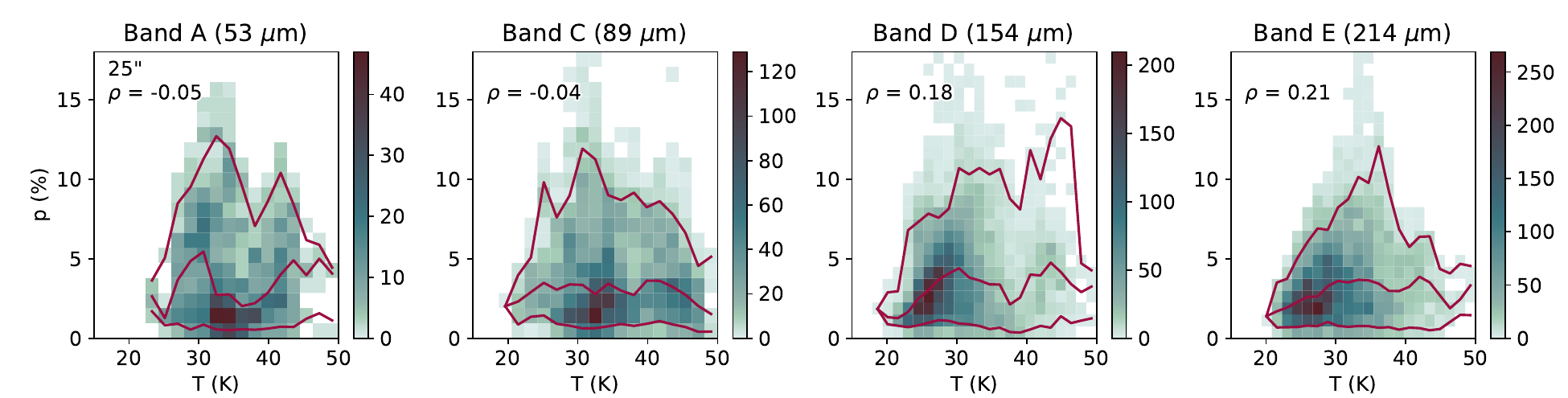}
    \caption{{2D histograms of $p$ as a function of $T$. The three red lines from bottom to top are the fifth percentile, median, and ninety-fifth percentile. These plots include data from all 26 regions at a resolution of $25''$. The calculated Spearman $\rho$ value is denoted on each plot.}}
    \label{fig:pvT}
\end{figure*}

In \autoref{fig:remaining_data}, we show the remaining HAWC+ maps with inferred magnetic field pseudovsctors that were not included in \autoref{fig:fam_port} or \autoref{fig:four_band}. \autoref{tab:obs_table} details information about each of the HAWC+ observations we use for out analyses. We include the available bands, the plan IDs, the original proposal PIs, the exposure time, the chop angles and amplitudes, the SOFIA Redux versions used to reduce the data, and the first publications to use the data. {In \autoref{fig:pvT}, we show 2D histograms of $p$ as a function of $T$ at each wavelength for all 26 regions at a resolution of $25''$.} 

\bibliography{Paper}{}
\bibliographystyle{aasjournal}

\begin{longtable*}
{P{1.8cm}|P{0.7cm}|P{2.35cm}|P{2.25cm}|P{1.2cm}|P{1.3cm}|P{1.15cm}|P{1.35cm}|P{3.1cm}}
\textbf{Region} & \textbf{Band} & \textbf{Plan ID} &\textbf{PI}&\textbf{Exposure (s)}&\textbf{Chop Angle ($^{\circ}$)} & \textbf{Chop Amp (")} &\textbf{Pipeline Version} &\textbf{Original Reference} \\
\endfirsthead
\toprule
L1527          &  D  &  70\_0609                                  &  GTO                         &   2251.9   &   -75    &  210  &  3.2.0       &  This Paper                           \\
L1688          &  A  &  70\_0609                                  &  GTO                         &   4730.6   &    30    &  240  &  3.2.0       &  This Paper                           \\
L1688          &  C  &  70\_0511, 70\_0609                      &  GTO                         &   3842.3   &    30    &  240  &  3.2.0       &  \citet{Santos2019}                   \\
L1688          &  D  &  70\_0511                                  &  GTO                         &    833.5  &    30    &  240  &  3.2.0       &  \citet{Santos2019}                   \\
$\rho$ Oph E/F        &  D  &  05\_0133,  06\_0116                     &  Giles Novak                                 &   2773.6    &   -72    &  240  &  3.2.0       &  \citet{Lee2021}, This Paper \\
L1689N         &  C  &  07\_0147&  Giles Novak&   1213.7   &   -85    &  200  &  3.2.0       &  \citet{encalada2024}                           \\
L1689N         &  D  &  06\_0116                                  &  Giles Novak                                 &    321.6   &   -85    &  210  &  3.2.0       &  \citet{encalada2024}                           \\
\hline\hline
NGC 1333        &  C  &  05\_0081, 06\_0099                      &  Fabio Santos                        &   5053.7   &  -108    &  240  &  3.2.0       &  This Paper                           \\
NGC 1333        &  E  &  06\_0098                                  &  Ian Stephens                                &   3595.9   &    90    &  250  &  3.2.0       &  \citet{Soam2024}                           \\
OMC1           &  A  &  70\_0609, 70\_0509                     &  GTO                         &   5762.3    &   125    &  230  &  1.3.0beta3  &  \citet{Chuss2019}                    \\
OMC1           &  C  &  88\_0005, 70\_0509                     &  GTO                         &   3697.7   &   -55    &  240  &  1.3.0beta3  &  \citet{Chuss2019}                    \\
OMC1           &  D  &  70\_0509                                  &  GTO                         &    756.8  &   125    &  230  &  1.3.0beta3  &  \citet{Chuss2019}                    \\
OMC1           &  E  &  70\_0509                                  &  GTO                        &   1034.8   &   125    &  230  &  1.3.0beta3  &  \citet{Chuss2019}                    \\
OMC2/3         &  D  &  07\_0026, 07\_0237                      & \raggedright{Stefan Heese, Joseph Michail}           &   3218.1   &   -90    &  250  &  2.7.0       &  \citet{Zielinski2022}, This Paper  \\
OMC2/3         &  E  &  07\_0026                                  &  Stefan Heese                                &   2876.9   &    90    &  250  &  3.2.0       &  \citet{Zielinski2022}                \\
NGC 2071       &  E  &  06\_0119                                  &  Ian Stephens                                &   2063.6   &    15    &  250  &  3.2.0       &  This Paper                           \\
NGC 2068       &  E  &  06\_0119                                  &  Ian Stephens                                &   1492.1   &    45    &  250  &  3.2.0       &  This Paper                           \\
Serpens  &  E  &  05\_0206, 06\_0088, 06\_0180, 07\_0238  &  \raggedright{Thushara Pillai, Paul Goldsmith}           &  10159.2    &   -50    &  150  &  3.2.0       &  \citet{Pillai2020}   \\
\hline
Vela C         &  C  &  70\_0609                                  &  GTO                         &   1661.3   &   -67.5  &  240  &  3.2.0       &  \citet{Bij2024}                      \\
Vela C         &  E  &  70\_0609                                  &  GTO                         &    946.7  &   -67.5  &  240  &  3.2.0       &  \citet{Bij2024}                      \\
S106           &  A  &  06\_0014                                  &  Dan Clemens                                 &    653.6  &   -60    &  250  &  2.7.0       &  This Paper                           \\
S106           &  E  &  06\_0014                                  &  Dan Clemens                                 &    474.6  &   -60    &  250  &  2.7.0       &  This Paper                           \\
NGC 6334        &  A  &  70\_0609                                  &  GTO                         &   1772.9   &  -135    &  210  &  3.2.0       &  This Paper                           \\
NGC 6334        &  C  &  70\_0609                                  &  GTO                         &    472.1  &  -135    &  210  &  3.2.0       &  This Paper                           \\
NGC 6334        &  D  &  70\_0609                                  &  GTO                         &    465.7  &  -135    &  210  &  3.2.0       &  This Paper                           \\
NGC 6334        &  E  &  70\_0609                                  &  GTO                         &    472.1  &  -135    &  210  &  3.2.0       &  This Paper                           \\
DR21           &  A  &  06\_0014, 07\_0013                      &  Dan Clemens                                 &   3902.9   &  -110    &  190  &  2.7.0       &  This Paper                           \\
DR21           &  E  &  06\_0014, 07\_0013                      &  Dan Clemens                                 &   3379.0   &  -110    &  250  &  2.7.0       &  This Paper                           \\
G34.43         &  E  &  06\_0039                                  &  Dan Clemens                                 &   1728.6   &  -110    &  200  &  3.2.0       &  This Paper                           \\
M17            &  A  &  70\_0609                                  &  GTO                         &    908.3   &  -100    &  225  &  3.2.0       &  This Paper                      \\
M17            &  C  &  70\_0609                                  &  GTO                         &   2332.6   &  -100    &  225  &  3.2.0       &  \citet{Cox2025}                      \\
M17            &  D  &  70\_0609                                  &  GTO                         &   1530.0   &     0    &  150  &  3.2.0       &  \citet{Hoang2022}                    \\
M17            &  E  &  70\_0609                                  &  GTO                         &    710.2  &  -100    &  225  &  3.2.0       &  \citet{Cox2025}                      \\
G14.2N         &  D  &  06\_0183, 07\_0066                      &  Fabio Santos                             &   1621.4   &   -95    &  240  &  2.7.0       &  This Paper                           \\
G14.2S         &  D  &  06\_0183                                  &  Fabio Santos                             &    892.09  &    40    &  240  &  3.2.0       &  This Paper                           \\
M16            &  C  &  06\_0059                                  &  Marc Pound                                  &   3367.7   &   -78    &  250  &  3.2.0       &  \citet{Sarkar2025}                           \\
M16            &  D  &  05\_0112                                  &  Marc Pound                                  &   1332.3   &   -82.7  &  250  &  3.2.0       &  \citet{Sarkar2025}                          \\
G351.77   &  E  &  07\_0238                                  &  Thushara Pillai                             &   1302.3   &   140    &  250  &  2.7.0       &  \citet{Jadhav2026}                           \\
W3OH           &  A  &  90\_0084, 70\_0509                      &  GTO  &   1000.7   &   -70    &  219  &  2.7.0       &  This Paper                           \\
W3OH           &  C  &  70\_0509                                  &  GTO                         &    919.1  &   -85    &  225  &  2.7.0       &  This Paper                           \\
W3OH           &  D  &  90\_0084                                  &  GTO                         &    446.0  &   -75    &  233  &  2.7.0       &  This Paper                           \\
W3OH           &  E  &  70\_0509                                  &  GTO                         &   1471.3   &   -85    &  225  &  2.7.0       &  This Paper                           \\
W3             &  A  &  90\_0084, 05\_0038                      &  \raggedright{GTO, John Vaillancourt}   &   2373.2   &   -70    &  219  &  2.7.0       &  This Paper                      \\
W3             &  C  &  90\_0084                                 &  GTO            &   1176.3   &   -70    &  219  &  2.7.0       &  \citet{Cox2025}                      \\
W3             &  D  &  88\_0005, 90\_0084                     &  GTO                         &   1681.8    &   -75    &  233  &  2.7.0       &  \citet{Cox2025}                      \\
W3             &  E  &  90\_0084                                  &  GTO                         &   2324.0   &   -75    &  233  &  2.7.0       &  \citet{Cox2025}                      \\
W33            &  A  &  05\_0038                                  &  John Vaillancourt                           &    839.0  &    35    &  250  &  3.2.0       &  This Paper                           \\
W33            &  C  &  05\_0038                                  &  John Vaillancourt                           &    306.4  &    30    &  219  &  3.2.0       &  This Paper                           \\
W33            &  E  &  05\_0038                                  &  John Vaillancourt                           &    422.9  &    25    &  250  &  3.2.0       &  This Paper                           \\
BYF77          &  D  &  07\_0089                                  &  Peter Barnes                                &   1117.8   &   -20    &  250  &  3.2.0       &  \citet{Barnes2025}                           \\
BYF73          &  D  &  07\_0089                                  &  Peter Barnes                                &    784.4  &   -65    &  250  &  3.2.0       &  \citet{Barnes2023}                   \\
Keyhole        &  C  &  07\_0081                                  &  Darren Dowell                               &  10303.2    &    -5    &  210  &  3.2.0       &  \citet{Seo2021}  \\

\caption{A table of every HAWC+ dataset used in our compilation. The last column lists the reference for the first publication of the data. For the datasets that list our paper and another paper, a portion of the data was published but we have included more observations. Entries in the Original Reference column that include a different paper and the text "This Paper" use a dataset that is a combination of previously published data and additional data from the archive.}
\label{tab:obs_table}
\end{longtable*}

\end{document}